\documentclass[pdftex,twocolumn,epjc3]{svjour3}          % twocolumn

\RequirePackage[T1]{fontenc}

\smartqed  % flush right qed marks, e.g. at end of proof

\RequirePackage{graphicx}
\RequirePackage{mathptmx}      % use Times fonts if available on your TeX system
\RequirePackage{flushend}
\RequirePackage[numbers,sort&compress]{natbib}
\RequirePackage[colorlinks,citecolor=blue,urlcolor=blue,linkcolor=blue]{hyperref}

\journalname{Eur. Phys. J. A}

\usepackage{amsfonts}
\usepackage{amssymb}
\usepackage{amsmath}
\usepackage{graphicx,color}
\usepackage{dcolumn}
\usepackage{epsfig}
\usepackage{bm}
\usepackage{bbm}
\usepackage{ulem,slashed,tensor,braket}

\usepackage{color}

\usepackage{hyperref}
\hypersetup{colorlinks=true,citecolor=blue,linkcolor=blue,urlcolor=blue}

\begin{document}

\title{Models of $\bm{J/\Psi}$ photo-production reactions on the nucleon}

%%%%%%%%%%%%%%%%%%%% Authors %%%%%%%%%%%%%%%%%%%%%%%%%%%%%%%%%

\author{T.-S. H. Lee\thanksref{e1,addr1} \and
S.~Sakinah\thanksref{e2,addr2} \and
Yongseok Oh\thanksref{e3,addr2,addr3}}

\thankstext{e1}{e-mail: tshlee@anl.gov}
\thankstext{e2}{ssakinahf@gmail.com}
\thankstext{e3}{e-mail: yohphy@knu.ac.kr}

\institute{Physics Division, Argonne National Laboratory, Argonne, Illinois 60439, USA\label{addr1} \and
Department of Physics, Kyungpook National University, Daegu 41566, Korea\label{addr2} \and
Asia Pacific Center for Theoretical Physics, Pohang, Gyeongbuk 37673, Korea\label{addr3}}

\date{Received: date / Accepted: date}

\maketitle

\begin{abstract}
The $J/\Psi$ photo-production reactions on the nucleon can provide information on 
the roles of gluons in determining the $J/\Psi$-nucleon ($J/\Psi$-N) interactions and
the structure of the nucleon. 
The information on the $J/\Psi$-N interactions is needed to test lattice QCD (LQCD) 
calculations and to understand the nucleon resonances such as $N^*(P_c)$ recently 
reported by the LHCb Collaboration. 
In addition, it is also needed to investigate the production of nuclei with hidden charms 
and to extract the gluon distributions in nuclei.  
The main purpose of this article is to  review six reaction models of $\gamma + p \rightarrow J/\Psi +p$ 
reactions which have been and can be applied to analyze the data from Thomas Jefferson 
National Accelerator Facility (JLab).
The formulae for each model are given and used to obtain the results to show the extent  
to which the available data can be described.
The models presented include the Pomeron-exchange model of Donnachie and
Landshoff ($Pom$-DL) and its extensions to include $J/\Psi$-N potentials extracted
from LQCD ($Pom$-pot) and to also use the constituent quark model (CQM) to account for
the quark substructure of $J/\Psi$ ($Pom$-CQM).
The other three models are developed from applying the perturbative QCD approach 
to calculate the two-gluon exchange using the generalized parton distribution (GPD) of 
the nucleon ($GPD$-based), two- and three-gluon exchanges using the parton distribution 
of the nucleon ($2g+3g$), and the exchanges of scalar ($0^{++}$) and tensor ($2^{++}$) 
glueballs within the holographic formulation ($holog$).
The results of investigating the excitation of the  nucleon resonances $N^*(P_c)$
in the  $\gamma + p \rightarrow J/\Psi +p$ reactions are also given.
We demonstrate that the differences between these six models can be unambiguously 
distinguished and the $N^*$ can be better studied by using the forthcoming JLab data 
at large $|t|$ and at energies very near the $J/\Psi$ production threshold.
Possible improvements of the considered models are discussed.
\end{abstract}

%\pacs{ 13.60.Le,  14.20.Gk}

%14.20.Gk Properties of specific particles, Baryon resonances (S=C=B=0)
%13.75.Gx Pion-baryon interactions

%13.60.Le Photon and charged-lepton interactions with hadrons; Meson production
%13.30.Eg Decays of baryons; Hadronic decays
%13.88.+e       Polarization in interactions and scattering
%11.80.La       Multiple scattering

\section{Introduction}

\begin{figure}[t]
\centering 
\includegraphics[width=0.9\columnwidth,angle=0]{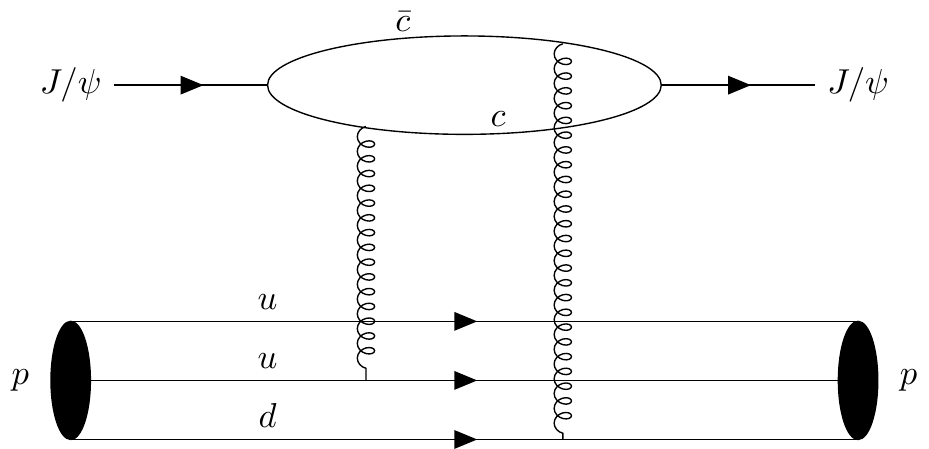}
\caption{One of the possible leading two-gluon exchange mechanisms of
$\gamma+ p \rightarrow J/\Psi+p$ reaction within QCD.}
\label{fig:fig-qcd}
\end{figure}

One of the important subjects in Quantum Chromodynamics (QCD) is  to understand 
the roles of gluons ($g$) in determining the structure of hadrons and 
hadron-hadron interactions. 
The progress in this direction can be made by investigating the interactions between the nucleon
and the quark-antiquark ($q\bar{q}$) systems which do not share the same up ($u$) and 
down ($d$) quarks with the nucleon.
The leading interaction between the nucleon (N) and the vector meson $J/\Psi$,
which is a charm-anticharm ($c\bar{c}$) system with spin-parity $J^\pi=1^-$,
is the two-gluon exchange mechanism, as illustrated in Fig.~\ref{fig:fig-qcd}.
Higher order multi-gluon exchange  effects can not be  neglected in the non-perturbative region.
The $J/\Psi$-N interaction can be estimated by using continuum and lattice studies at low energies
and the heavy quark effective field theory and perturbative QCD (PQCD) at high energies.
Alternatively, one can extract it from the data of $J/\Psi$  photo-production reactions within 
appropriate reaction models. 
The information on the $J/\Psi$-N interaction is needed to understand the nucleon
resonances $N^*(P_c)$  reported by the LHCb 
collaboration~\cite{LHCb-15,LHCb-16a,LHCb-19,LHCb-21a}.
It is also needed to extract the gluonic distributions in nuclei, and to study the 
existence of nuclei with hidden charms~\cite{BD88a,BSD90,GLM00,BSFS06,WL12}.

The data of  $\gamma + p \rightarrow J/\Psi+ p $ reactions will be extensive and precise
from Jefferson Laboratory (JLab) with 12 GeV upgrade in the near 
future~\cite{GlueX-19,MJPC16,JLab-16,ATHENNA-12}.
The  purpose of this paper is to review six reaction models which have been applied to 
investigate the first published JLab data~\cite{GlueX-19}, and can be used to analyze the forthcoming data.  
We will present formulas which are used to obtain the results presented in this paper for 
examining the extent to which the available data can be described by each model.
Predictions will be presented for future experimental tests.
In this section we will briefly describe the essential ingredients of these six models. 
The details are then given in the rest of the paper.

\begin{figure}[t]
\centering
\includegraphics[width=0.9\columnwidth,angle=0]{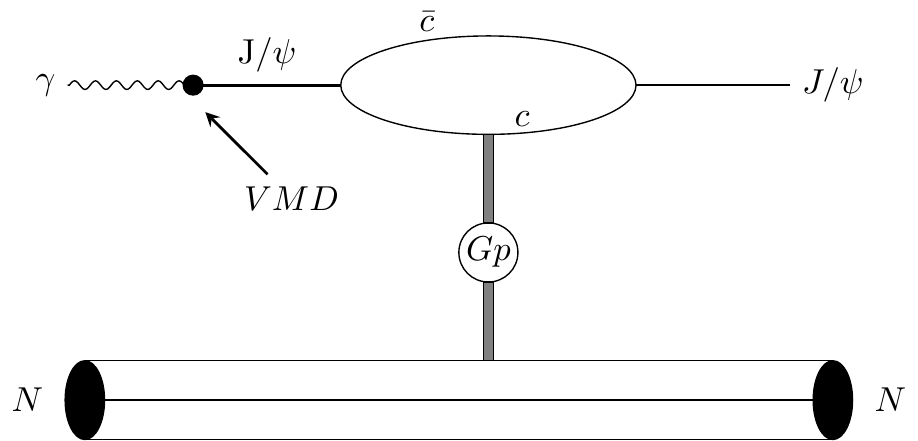} 
\vskip 1cm
\includegraphics[width=0.9\columnwidth,angle=0]{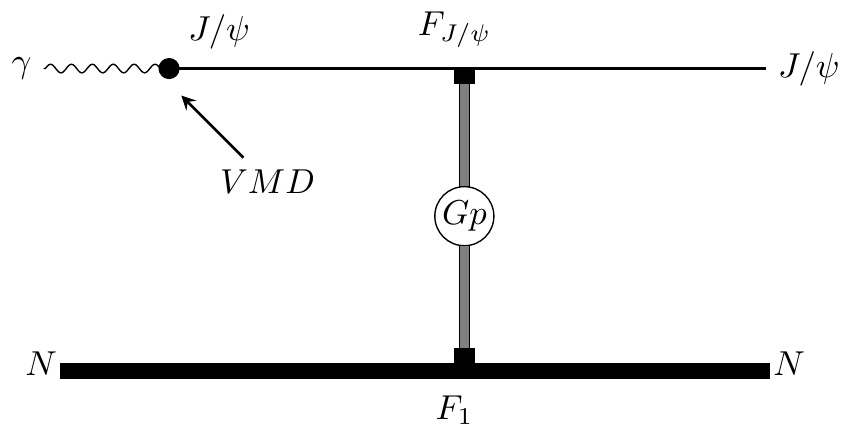}
\caption{ 
Pomeron-exchange model of Donnachie and Landshoff ($Pom-DL$).
Upper: Pomeron-exchange between quarks in $J/\Psi$ and nucleon,
Lower: Pomeron-exchange amplitude Eq.~(\ref{eq:amp-pom}) resulted from 
assuming the Pomeron-photon analogy and using the factorization approximation.}
\label{fig:fig-pom-dl}
\end{figure}

The study of photo-production of vector mesons ($V$) at high energies has a long 
history~\cite{Collins,IW77,DDLN} with extensive data sets~\cite{ZEUS-95b,BDLM79}.
A major advance was made by Donnachie and Landshoff~\cite{DL84,DL92,DL95,DL98} (DL) who
demonstrated that the data of $\gamma + p \rightarrow V +p$ reactions at high energies 
can be described very well by using the Vector Meson Dominance (VMD) assumption~\cite{Sakurai60,GZ61,Sakurai69,SS72a} 
and the Pomeron-exchange mechanism within the Regge phenomenology~\cite{Collins,IW77}.  
The connection of the DL model to QCD was qualitatively justified in some investigations~\cite{Low75,Nussinov75,DL88}
which attempted to relate the Pomeron-exchange to  the gluon-exchange. 
In this approach, the incoming photon is converted into a vector meson $V$ which is then scattered  
from the nucleon by the Pomeron-exchange between quarks in two hadrons, as illustrated in  
the upper part of Fig.~\ref{fig:fig-pom-dl}.
The main assumption of the DL  model is the Pomeron-photon analogy~\cite{LP71a,JL74}
that Pomerons interact with quarks like $C=+$ isoscalar photons and the usual factorization 
approximation can be used to simplify the loop-integration over quark wavefunctions of hadrons. 
Thus the emission or absorption of a Pomeron by a hadron can be calculated by using the 
electromagnetic form factors of hadrons, such as $F_1(t)$ of the nucleon.
As illustrated in the lower part of Fig.~\ref{fig:fig-pom-dl}, the amplitudes for $\gamma +p \rightarrow V +p$ 
within the DL model can be written schematically as
\begin{eqnarray}
t^{\rm Pom} \sim \left[ \frac{e}{f_{V}}F_{V}(t) \right] \times G_P(s,t)\times F_1(t) ,
\label{eq:amp-pom}
\end{eqnarray}
where $f_V$ is determined by the width of $V\rightarrow e^+e^-$ decay using the VMD assumption, 
$s$ and $t$  are the usual Mandelstam variables, $F_{V}(t)$ is the form factor for $V$, and 
$G_{P}(t,s)$ is the Pomeron propagator of the Regge phenomenology.
The model defined by Eq.~(\ref{eq:amp-pom}) will be called $Pom$-DL model in this paper.

With the parameters determined by fitting~\cite{WL12,WLZ19,Lee20,KLNO21} the total cross section 
data of Refs.~\cite{ZEUS-95b, H1-96,GHLL75,CLPS75} of $\gamma+ p\rightarrow J/\Psi+ p$ up to invariant 
mass $W=300$~GeV, it was found~\cite{Lee20} that the $Pom$-DL model can not describe the JLab 
data at $W <  7$~GeV. 
This is not unexpected because the $J/\Psi$-N interactions calculated by using the Pomeron-photon analogy 
is only part of possible gluon-exchange mechanisms.
In the near threshold energy region where the relative velocity  of the outgoing $J/\Psi$-N is not large and
the higher order multi-gluon exchanges can not be neglected.
In addition, the $J/\Psi + N \rightarrow J/\Psi +N $ transition amplitude $t_{J/\Psi N, J/\Psi N}(k,q,W)$ 
near threshold is far off-shell; for example, at $W= (m_N+m_{J/\Psi})+ 0.5 $~GeV, the incoming $\gamma N$ 
relative momentum is $q= 0.8$~GeV which is much larger than the outgoing $J/\Psi$-N relative momentum 
$k=0.1$~GeV in the CM system.
Hence the Regge phenomenology, which is based on the on-shell formulation, is not directly applicable for 
describing $J/\Psi+N \rightarrow J/\Psi+N$ in the near threshold energy region.

\begin{figure}[t]
\centering
\includegraphics[width=0.9\columnwidth,angle=0]{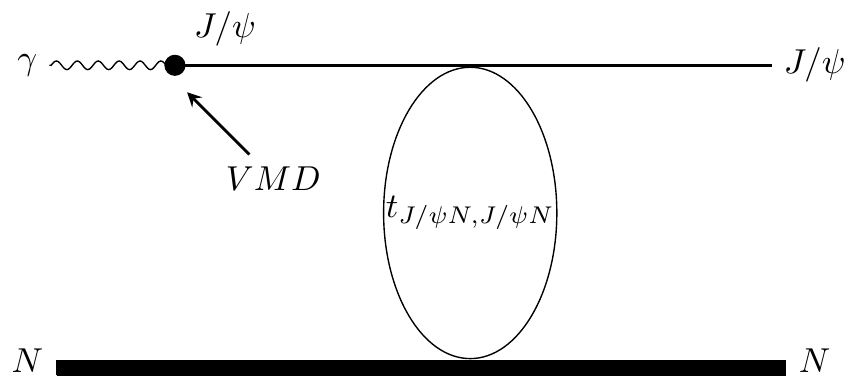}
\vskip 1cm
\includegraphics[width=0.9\columnwidth,angle=0]{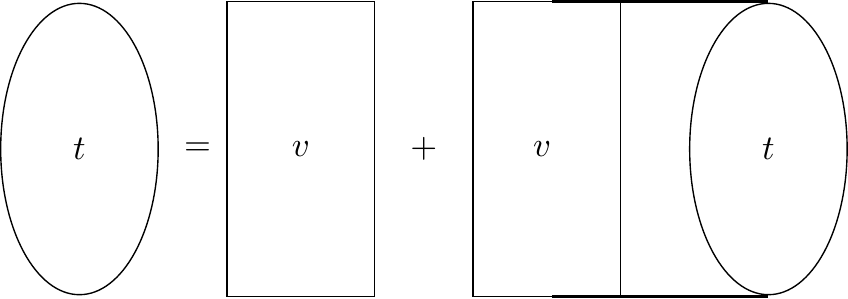}
\caption{$Pom$-pot model. Upper: The amplitude of Eq.~(\ref{eq:amp-pot}).
Lower: $J/\Psi$-N scattering equation~(\ref{eq:amp-pot-1}). 
Here, $v$ and $t$ stand for $v_{J/\psi  N,J/\Psi N}$ and $t_{J/\psi  N,J/\Psi N}$, respectively.}
\label{fig:fig-pom-pot}
\end{figure}

An obvious next step~\cite{Lee20} to improve the $Pom$-DL model is to add a $J/\Psi$-N scattering 
amplitude generated from a $J/\Psi$-N potential $v_{J/\Psi N,J/\Psi N}$ which can be interpreted as the 
multi-gluon exchange mechanisms.
By also using the VMD assumption, it gives the amplitude, illustrated in the upper part of Fig.~\ref{fig:fig-pom-pot},  
which is of the following form,
\begin{eqnarray}
t^{\rm pot}=\frac{e\,m^2_{J/\Psi}}{f_{J/\Psi}}
\frac{1}{q^2-m^2_{J/\Psi}}t_{J/\Psi N,J/\Psi N} ,
\label{eq:amp-pot}
\end{eqnarray}
where $m_{J/\Psi}$ is the mass of $J/\Psi$, and the $J/\Psi$-N scattering amplitude is defined by
the following Lippmann-Schwinger equation:
\begin{eqnarray}
t_{J/\Psi N,J/\Psi N} &=& v_{J/\Psi N,J/\Psi N}
\nonumber \\ && \mbox{}
+ v_{J/\Psi N,J/\Psi N} \, G_{J/\Psi N}(W) \, t_{J/\Psi N,J/\Psi N}.
\label{eq:amp-pot-1}
\end{eqnarray}
Here, $G_{J/\Psi N}(W)$ is the propagator of the $J/\Psi$-N system. 
Eq.~(\ref{eq:amp-pot-1}) is illustrated in the lower part of Fig.~\ref{fig:fig-pom-pot}. 
Adding $t^{\rm pot}$ to the $Pom$-DL amplitude, we then obtain a $Pom$-pot model defined by the following 
amplitude
\begin{eqnarray}
t^{\rm Pom+pot}= t^{\rm Pom} + t^{\rm pot} .
\label{eq:amp-pom-pot}
\end{eqnarray}

The results from $Pom$-pot model depend on the choice of $v_{J/\Psi N,J/\Psi N}$ in solving Eq.~(\ref{eq:amp-pot-1}).
Several attempts had been made to determine $v_{J/\Psi N,J/\Psi N}$.
An important step was taken by Peskin~\cite{Peskin79} who applied the operator product expansion to 
evaluate the strength of the color field emitted by heavy $q\bar{q}$ systems, and suggested~\cite{BP79} 
that the van der Waals force induced by the color field of $J/\Psi$ on nucleons can generate an attractive short-range 
$J/\Psi$-N interaction. 
The results of Peskin were used by Luke, Manohar, and Savage~\cite{LMS92} to predict, using the effective field theory 
method, the $J/\Psi$-N forward scattering amplitude which was then used to get an estimation that $J/\Psi$ can have 
a few MeV/nucleon attraction in nuclear matter.
The $J/\Psi$-N forward scattering amplitude of Ref.~\cite{LMS92} was further investigated by Brodsky and Miller~\cite{BM97a}
to derive a $J/\Psi$-N potential which gives a $J/\Psi$-N scattering length of $-0.24$~fm.
The result of Peskin was also used by Kaidalov and Volkovitsky~\cite{KV92}, who differed from Ref.~\cite{BM97a} in 
evaluating the gluon content in the nucleon, to give a much smaller scattering length of $-0.05$~fm.

In LQCD calculations using the approach of Refs.~\cite{IAH06,AHI09}, Kawanai and Sasaki~\cite{KS10b,KS11} 
obtained an attractive $J/\Psi$-N potential of the Yukawa form $v_{J/\Psi N,J/\Psi N} = - \alpha  e^{-\mu r}/{r}$
with $\alpha =  0.1$ and $\mu = 0.6$~GeV, which gives a scattering length of $- 0.09$~fm. 
Using the potential $v_{J/\Psi N,J/\Psi N}(r)$ extracted from this LQCD calculation, the $Pom$-pot model was used in 
Ref.~\cite{Lee20} to fit the JLab data by including a reduction of the VMD coupling constant $1/f_{J/\Psi}$ by a factor 
of about 0.4--0.7.  
The need of such a correction on the VMD coupling constant is consistent with the earlier study of $\rho$ photo-production~\cite{DL95}. 
It is also justified since the VMD coupling constant is determined by the $J/\Psi \to \gamma \to  e^+e^-$ decay width
at $q^2=m^2_{J/\Psi} \sim 9~\mbox{GeV}^2$ which is far from $q^2=0$ of the $\gamma + p \to J/\Psi+p$ reaction. 
Furthermore, the use of VMD for $J/\Psi$ is questionable as discussed in Ref.~\cite{XCYBCR21}.
Thus, the phenomenological procedure of adjuting the off-shell factor $1/f_{J/\Psi}$ 
makes the $Pom$-pot model uncertain in testing the LQCD calculations of $J/\Psi$-N potentials.
Nevertheless, the $Pom$-pot model is the only model which fits the data from threshold to very high energy $W=300$~GeV and 
therefore can be applied to use the data of $J/\Psi$ photo-production on nuclei to  study the gluonic distribution in nuclei 
and the existence of nuclei with hidden charms~\cite{BD88a,BSD90,GLM00,BSFS06,WL12}.

\begin{figure}[t]
\centering
\includegraphics[width=0.9\columnwidth,angle=0]{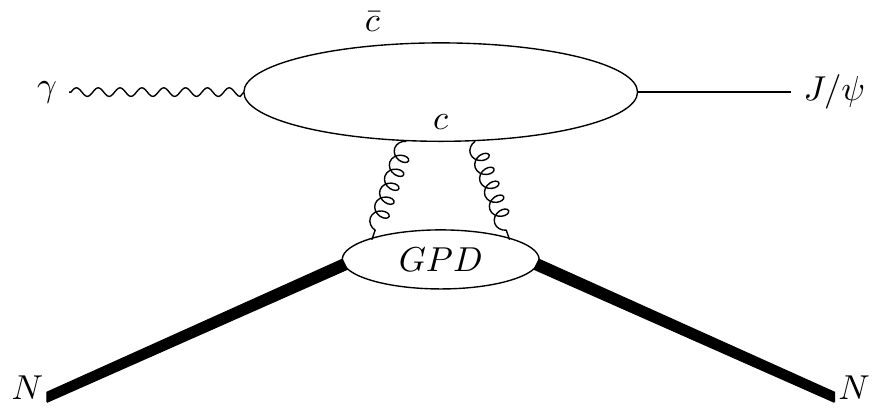}
\vskip 1cm
\includegraphics[width=0.9\columnwidth,angle=0]{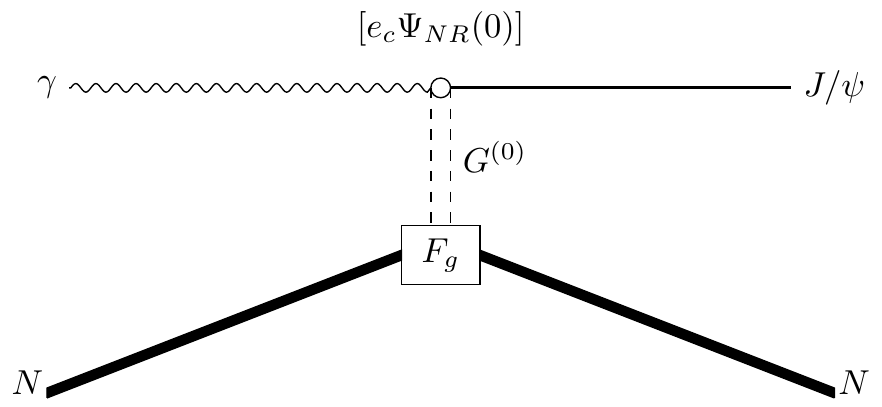}
\caption{    
$GPD$-based model. Upper: One of the four two-gluon exchange diagrams of Eq.~(\ref{eq:gpd-exact}), 
Lower: The amplitude of Eq.~(\ref{eq:crst-gpd}). }
\label{fig:gpd}
\end{figure}

The models described above are for analyzing the data from threshold to very high energies up to $W=300$~GeV. 
Focusing on the near threshold region $W \leq 7$~GeV,  two models based on PQCD
have  been developed to describe the JLab data.
Both models assume that the incoming photon fluctuates into a $c\bar{c}$ pair which is then scattered from  
the nucleon by gluon-exchange mechanisms.

The model of Ref.~\cite{GJL21} starts with the two-gluon-exchange amplitude which is illustrated in the upper part of
Fig.~\ref{fig:gpd} and has the following form~\cite{guo-1}
\begin{eqnarray}
&& t_{J/\Psi N,\gamma N}(P' K; P q) 
\nonumber \\
&=&
 (ig)^2  \int \frac{d^4k \, d^4 l}{(2\pi)^8} 
\braket{ P' \mid A^\mu(-l- \textstyle\frac{\Delta}{2}) A^\nu(l-\textstyle\frac{\Delta}{2})|P } 
\nonumber \\ && \times 
\mbox{Tr} \left[ \sum_{n=a}^{d}I^n_{\mu\nu}(K,k,l,\Delta,\epsilon) \Psi(K,k) \right] ,
\label{eq:gpd-exact}
\end{eqnarray}
where $P$ and $P'$ are the momenta of the initial and final nucleons, $q$ the incident photon momentum,  
$\epsilon$ the photon polarization vector, $K$ the final $J/\Psi$ momentum, $\Delta \equiv P'-P$, 
$A^\mu$ the gluon field, $\Psi(K,k)$ the wavefunction of $c\bar{c}$ in $J/\Psi$, and $I^a_{\mu\nu}(K,k,l,\Delta,\epsilon)$
the propagators of quarks in $J/\Psi$. 
The above expression can be simplified by using the heavy quark expansion in $1/m_{J/\Psi}$ and the non-relativistic 
limit of $\Psi(K,k)$. 
The  amplitude is then reduced into a form related to the moment expansion of Generalized Parton Distribution (GPD). 
By keeping only the $n=0$ moment, the cross section is then determined by a gluonic form factor of the nucleon and 
the non-relativistic wavefunction $\phi_{NR}(0)$ of $J/\Psi$ at relative distance $r=0$ of two quarks.
The resulting amplitude is illustrated  in the lower part of Fig.~\ref{fig:gpd}. 
The  differential cross section can then be written as
\begin{eqnarray}
\frac{d\sigma}{dt}=\frac{\alpha_{EM}e^2_c}{4(W^2-m^2_N)^2}\frac{(16\pi\alpha_S)^2}{3m^3_{J/\Psi}}
\left| \phi_{NR}(0) \right|^2 |G^{(0)}(t,\xi)|^2 ,
\label{eq:crst-gpd}
\end{eqnarray}
where $\alpha_{EM}=e^2/(4\pi)=1/137$, $e_c=\frac{2}{3}e$,  $\alpha_S$ is a QCD coupling constant, and $G^{(0)}(t,\xi)$ 
is the $n=0$ moment of GPD defined  by 
\begin{equation}
G^{(0)}(t,\xi) =\frac{1}{\xi^{2}} \int_{-1}^{+1} dx\,F_g(x,\xi,t) .
\label{eq:gpd-n0}
\end{equation}
Here the skewness $\xi$ is defined by $\xi=\frac{P^+-P^{'+}}{P^++P^{'+}}$ with $P^+=\frac{1}{2}(P^0+P^3)$, and
$F_g(x,\xi,t)$ is the gluon GPD of the nucleon.
The parameters of $G^{(0)}(t,\xi)$ had been determined~\cite{SD18} by using LQCD, but were adjusted to fit the JLab data.
The model defined by Eqs.~(\ref{eq:crst-gpd}) and (\ref{eq:gpd-n0}) will be called $GPD$-based model in this paper.

\begin{figure}[t]
\centering
\includegraphics[width=0.9\columnwidth,angle=0]{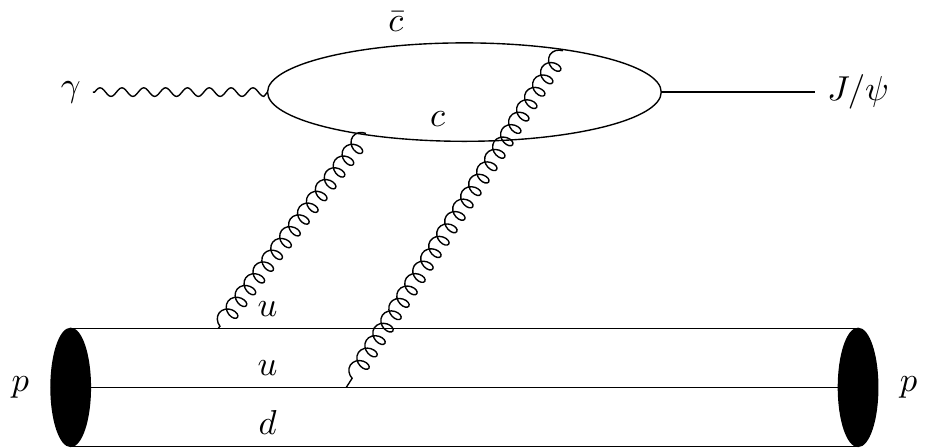}
\vskip 1cm
\includegraphics[width=0.9\columnwidth,angle=0]{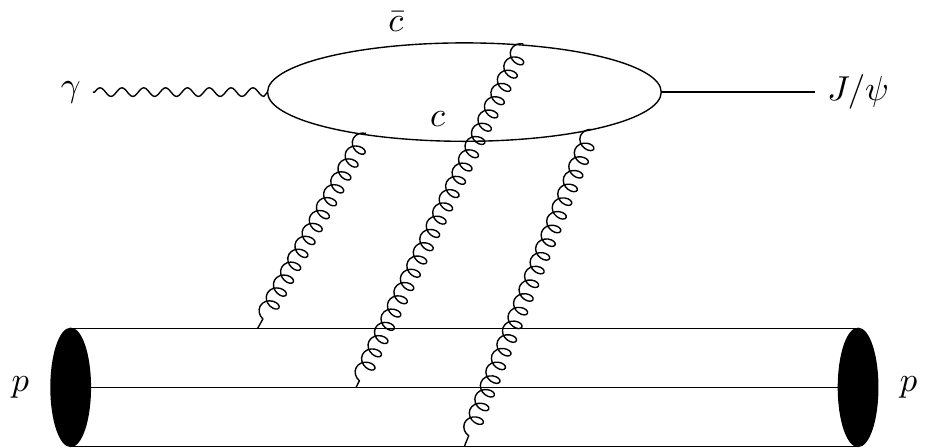}
\caption{The $2g+3g$ model. Upper: two-gluon exchange, Lower: three-gluon exchange.}
\label{fig:2g3g}
\end{figure}

The PQCD model of Ref.~\cite{BCHL01}, applied in Ref.~\cite{GlueX-19} to fit the JLab data, assumes that the $c\bar{c}$ 
scattered from the valance quarks with fraction of momentum $x\sim 1$ of  the target nucleus.
For the nucleon target, this model has two- and three-gluon exchange processes, as illustrated in Fig.~\ref{fig:2g3g}.
As discussed in Ref.~\cite{BBS94}, this implies that the production rate behaves near $x \to 1$ as $(1-x)^{2n_s}$, where 
$n_s$ is the number of spectators.
They also follow Ref.~\cite{BB79} to assume that the probability that a quark in the proton of radius $R$ is found within
the transverse distance $1/M_{c\bar{c}}$ of $c\bar{c}$ from proton is $1/(R^2M^2_{c\bar{c}})$, where $M_{c\bar{c}}$ is  
the mass of $c\bar{c}$.
For exclusive $J/\Psi$ production, a form factor $F_{ng}(t)$  is needed to describe how the scattered quarks combine with 
the spectator quarks to form a proton. 
They then obtain
\begin{eqnarray}
\frac{d\sigma_{2g}}{dt}=N_{2g}\frac{(1-x)^2}{R^2M^2_{c\bar{c}}}F^2_{2g}(t)(s-m_p^2) ,  
\label{eq:2g}\\
\frac{d\sigma_{3g}}{dt}=N_{3g}\frac{(1-x)^0}{R^4M^4_{c\bar{c}}}F^2_{3g}(t)(s-m_p^2) ,
\label{eq:3g}
\end{eqnarray}
where  the factor $(s-m^2_p)$ is from the coupling  of photon to $ c\bar{c}$.
It is rather uncertain to estimate $x$ and they assumed
\begin{eqnarray}
x=(2m_pM_{c\bar{c}}+M_{c\bar{c}}^2)/(s-m^2_p) ,
\end{eqnarray}
where  $s=E^2_{cm}$ with $E_{cm}$ being  the total energy in the center of mass frame.
The form factors $F_{2g}(t)$ and $F_{3g}(t)$ are unknown and must be determined from experiments. 
Using rather scarce data of $t$-dependence of $d\sigma/dt$ near threshold, they set $F^2_{2g}(t)=F^2_{3g}(t)=\exp(1.13t)$,
where $t$ is in the units of GeV$^2$.
The constants $N_{2g}$ and $N_{3g}$ are then determined from fitting the total cross section data of $\gamma + p\rightarrow J/\Psi  +p$  
reactions.
The model defined by Eqs.~(\ref{eq:2g}) and (\ref{eq:3g}) will be called $2g+3g$ model in this paper.

\begin{figure}[t]
\centering
\includegraphics[width=0.9\columnwidth,angle=0]{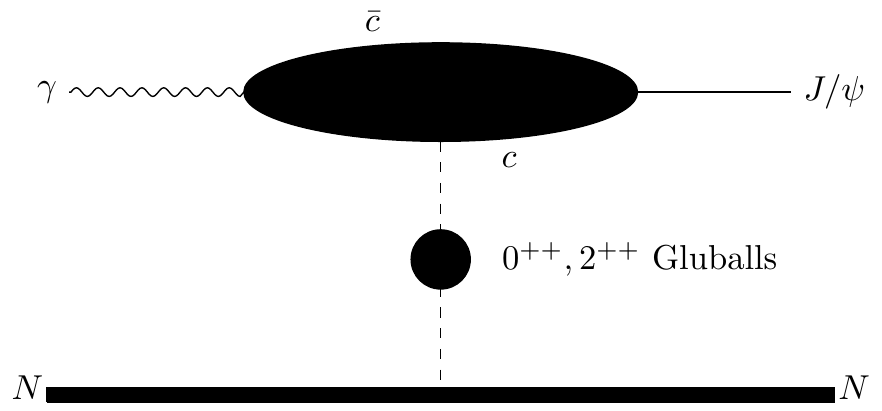}
\caption{ The holographic model. }
\label{fig:holo}
\end{figure}

We will also give formula of the model of Refs.~\cite{MZ19,MZ21} based on the holographic QCD.
This model assumes that the $J/\Psi$ photo-production is due to the exchanges of tensor ($2^{++}$) and scalar ($0^{++}$)
glueballs, as illustrated in Fig.~\ref{fig:holo}. 
This model will be called $Holog$ model in this paper.

\begin{figure}[t]
\centering
\includegraphics[width=0.9\columnwidth,angle=0]{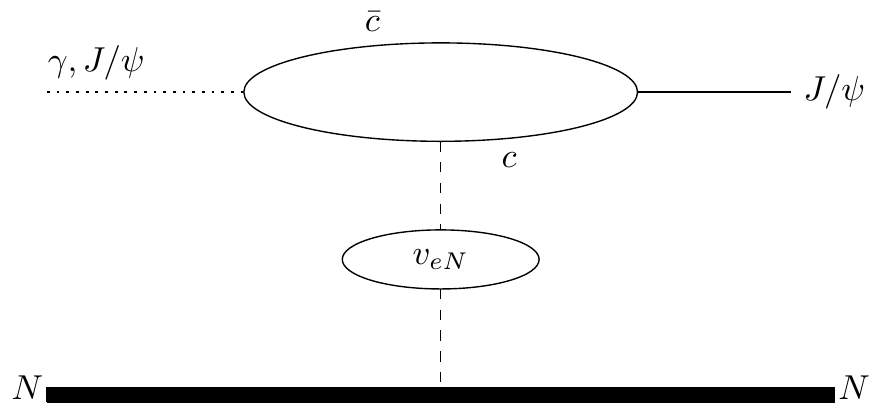}
\vskip 1cm
\includegraphics[width=0.9\columnwidth,angle=0]{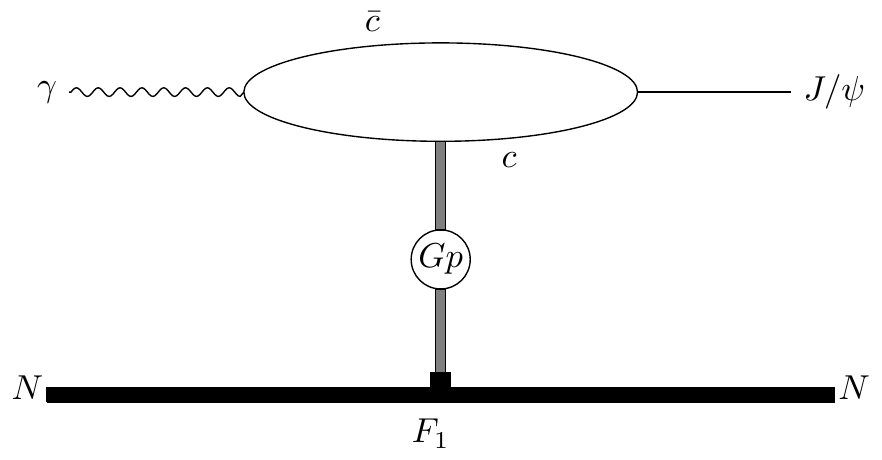}
\caption{Models with $c\bar{c}$-loop mechanisms. 
Upper:  calculated from quark-nucleon potential ($v_{cN}$), 
Lower:  calculated from Pomeron-exchange mechanism. }
\label{fig:fig-pom-dl-loop}
\end{figure}

The sixth model we will present is aimed at testing the  $J/\Psi$-N potentials extracted from LQCD calculations.
It is obtained by removing the VMD assumption, which is questionable~\cite{XCYBCR21} for  $J/\Psi$, in the $Pom$-pot model 
described above.
If we further assume that the $c\bar{c}$-N interaction can be defined by a quark-N potential $v_{cN}$, then the $\gamma+ N 
\rightarrow J/\Psi+N$ and $J/\Psi+N\rightarrow J/\Psi+N$ amplitudes in $t^{\rm pot}$ of 
Eqs.~(\ref{eq:amp-pot}) and (\ref{eq:amp-pot-1}) 
are defined by $c\bar{c}$-loop mechanisms, as illustrated in the upper part of Fig.~\ref{fig:fig-pom-dl-loop}. 
Similarly, the Pomeron-exchange term $t^{\rm Pom}$ should also be defined by the same $c\bar{c}$-loop mechanism, as illustrated 
in the lower part of Fig.~\ref{fig:fig-pom-dl-loop}. 
By using a hadron model~\cite{RW94,Roberts94} based on Dyson-Schwinger equation (DSE) of QCD, such a quark-loop
Pomeron-exchange model was explored in Refs.~\cite{PL96,PL97}. 
It will be interesting to use the recent DSE models~\cite{MR97a,CLR10,CCRST13,QR20,YBCR22,AV00,SEVA11,EF11} 
to improve the results of Refs.~\cite{PL96,PL97} and to also evaluate loop mechanisms with $v_{cN}$ 
(upper part of Fig.~\ref{fig:fig-pom-dl-loop}) which is needed to describe the data near threshold.

Alternatively, one can use the constituent quark model (CQM)~\cite{IK79,Richard12,SEFH13} to evaluate the $c\bar{c}$-loops
of Fig.~\ref{fig:fig-pom-dl-loop}.
This approach is simpler in practice and is reasonable since CQM for the heavy quark systems, such as $J/\Psi$ and $\Upsilon$, 
has been well established~\cite{Richard12}.
In this paper, we will detail how such a CQM-based model, referred to as $Pom$-CQM model, can relate the $J/\Psi$ potentials 
extracted from LQCD calculations to the $J/\Psi$ photo-production data.

The $N^*$ excitation in the $J/\Psi$ photo-production has been studied in Ref.~\cite{WLZ19} within the dynamical formulation 
developed in Refs.~\cite{SL96,MSL06,JLMS07,KNLS13}.
In this Argonne-Osaka approach, the total amplitude including the excitation of $N^*$ is of the following form:
\begin{eqnarray}
T_{VN,\gamma N}(W)= t_{VN,\gamma N}(W) + t^{N^*}_{VN,\gamma N}(W) ,
\label{eq:nstar-1}
\end{eqnarray}
where $t_{VN,\gamma N}$ can be generated from any of the models described above, and
\begin{eqnarray}
t^{N^*}_{VN, \gamma N}(W)&=& 
\bar{F}^+_{N^*,VN}(W)\frac{1}{W- M^*_0 -\Sigma(W)}\bar{F}_{N^*,\gamma N}(W) ,
\label{eq:nstar-2}
\end{eqnarray}
where $\bar{F}_{N^*,MB}$ is the dressed $N^*\rightarrow MB$ vertex function, and $\Sigma(W)$ is the self-energy of the 
resonance with a bare mass $M^*_0$.
The $VN \rightarrow VN$ scattering amplitudes are included in $\bar{F}_{N^*,VN}$ and $\Sigma(W)$, as required by 
the unitarity condition.
The  above dynamical formulation can be used to investigate whether $N^*(P_c)$ states are the meson-baryon 
molecules~\cite{SGXZ16,RNO15,MO15,LSGZ17} or the compact pentaquark states~\cite{MPR15,LHH15,Wang15a}.
We will present results from using Eqs.~(\ref{eq:nstar-1}) and (\ref{eq:nstar-2}) to investigate the $N^*(P_b)$ reported 
in Ref.~\cite{LHCb-21a}.

In Sec.~\ref{sec:Pom}, we briefly review the Regge phenomenology for explaining the Pomeron-exchange model 
of Donnachie and Landshoff ($Pom$-DL).
The $Pom$-pot model obtained from extending the $Pom$-DL model to include $J/\Psi$-N potentials extracted from 
LQCD is presented in Sec.~\ref{sec:Pom+pot}.
The models based on GPD ($GPD$-based), two-gluon and three-gluon exchange ($2g+3g$), and holographic 
approach ($Holog$) are  presented in Sec.~\ref{sec:pqcd}. The study of $N^*$ excitation is given in Sec.~\ref{sec:N*}. 
Section~\ref{sec:pom-cqm} is for presenting the $Pom$-CQM model.
Predictions for future experimental tests are given in Sec.~\ref{sec:predict}.
In Sec.~\ref{sec:summary}, we give a summary and discuss possible future developments.

\section{Pomeron-exchange model} 
\label{sec:Pom}

The Pomeron-exchange model of photo-production of vector mesons was developed within the Regge phenomenology.
As can be seen from extensive literature~\cite{Collins,IW77,DDLN}, the Regge phenomenology cannot be
derived rigorously from relativistic quantum field theory, although some understanding of Regge poles have been 
obtained in Ref.~\cite{LS62a}.
It was largely from the study~\cite{Regge59,Regge60,BLR62} of potential scattering within the non-relativistic
quantum mechanics and was simply extended to relativistic formulation~\cite{ Chew62a,CF62a,Mandelstam63a} 
of scattering amplitudes.
For our purposes, we give in \ref{sec:appendix} sufficiently self-contained explanations which are needed to develop the
formulas of Pomeron-exchange models.

With the derivations given in \ref{sec:appendix}, we start with Eq.~(\ref{eq:regge-f}) for the amplitude $T(s,t)$ of the process  
$1(p_1)+2(p_2) \rightarrow 3(p_3)+4(p_4)$.
By defining the usual Mandelstam variables as
\begin{eqnarray}
s&=&(p_1+p_2)^2 ,\\
t&=&(p_1-p_3)^2 ,
\end{eqnarray}
we  have
\begin{eqnarray}
T(s,t)&=&\sum_{n} \beta^{13}_n(t) \beta^{24}_n(t)
\frac{1+s_n e^{-i\pi\alpha_n(t)}}{2\sin[\pi \alpha_n(t)]}(\alpha_{1,n}s)^{\alpha_n(t)} ,
\label{eq:regge-f0}
\end{eqnarray}
where $s_n=+1$ $(-1)$ corresponds to even (odd) parity exchanges, $\beta^{13}_n(t)$ and $\beta^{24}_n(t)$ characterize 
the hadron structure, and $\alpha_n(t)=\alpha_{0,n}+\alpha_{1,n}t$ defines the Regge trajectory $n$.
We now note that the amplitude $T(s,t)$ has singularities in $t$ defined by $\alpha_n(t)= \mbox{integers}$.
It has the feature of the one-particle-exchange amplitude in the relativistic quantum field theory,
\begin{eqnarray}
T(s,t) \sim \frac{1}{t-m^2} ,
\end{eqnarray}
which has singularity at $t=m^2$. 
Thus the amplitude~(\ref{eq:regge-f0}) can be interpreted as the exchange  of particles with masses $M_{L_n}$ defined
by $\alpha_n(t=M^2_{L_n})=  L_n$ with $L_n$ being the spin quantum number of the exchanged particle.
This is an intuitively very attractive interpretation of the scattering amplitude~(\ref{eq:regge-f0}). 
However, there exists no successful derivation of Eq.~(\ref{eq:regge-f0}) from relativistic quantum field theory and the 
form factors $\beta^{13}_n(t)$ and $ \beta^{24}_n(t)$ are determined experimentally or calculated from hadron models.

We now turn  to explaining how the Pomeron-exchange is introduced in Regge phenomenology.
It was from examining the total cross sections $\sigma^{\rm tot}$ defined by the amplitude of Eq.~(\ref{eq:regge-f0}):
\begin{eqnarray}
\sigma^{\rm tot}&=&\int\,dt\,\frac{d\sigma}{dt} \nonumber \\
 &=& \int  \,dt \left( \frac{1}{16\pi s^2} \right)  |T(s,t)|^2 \,.
\label{eq:totcrst} 
\end{eqnarray}

\subsection{Hadron-hadron scattering}

\begin{figure}[t] 
\centering
\includegraphics[width=0.9\columnwidth,angle=0]{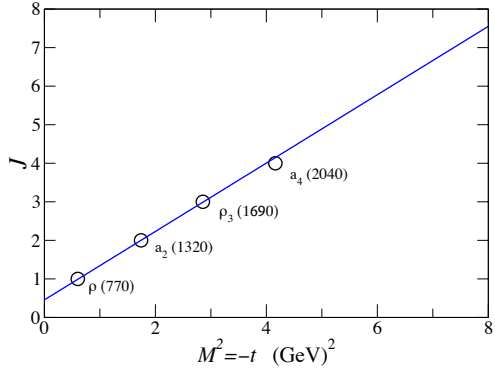}
\caption{
The $\rho$ trajectory determined by the masses of $\rho(760)$, $\mathrm{a}_2(1320)$, $\rho_3(1690)$ and $\mathrm{a}_4(2040)$.  }
\label{fig:rho-traj}
\end{figure}

The approach based on Eqs.~(\ref{eq:regge-f0}) and (\ref{eq:totcrst}) had been applied~\cite{DL79,DL83a,DL84a} to investigate 
the hadron-hadron scattering. 
The first step is to use the existing particle spectra to determine the Regge trajectories.
It was found that the trajectories for $n =\pi, \rho, \dots$ can be written as 
\begin{eqnarray}
\alpha_n(t)=\alpha_{0,n} + \alpha_{1,n}t ,
\end{eqnarray}
where
\begin{eqnarray}
\alpha_{0,n} < 1 .
\label{eq:alpha0}
\end{eqnarray}
An example for the $\rho$ trajectory is illustrated in Fig.~\ref{fig:rho-traj} with $\alpha_0 = 0.456$ and $\alpha_1 = 0.887$~GeV$^{-2}$.
If these trajectories are included in calculating the total cross section~(\ref{eq:totcrst}), we then find that at very large $s$, 
only one trajectory $n_{s}$ dominates. 
Since  $\alpha_{0,n}<1$, Eq.~(\ref{eq:totcrst}) leads to
\begin{eqnarray}
\sigma^{\rm tot}(s) \sim s^{2(\alpha_{0,n_s}-1)} =  s^{-m} \,;\quad m > 0.
\label{eq:totcrst-1}
\end{eqnarray}
Eq.~(\ref{eq:totcrst-1}) contradicts with the experimental data that the total cross sections $\sigma^{\rm tot}(s)$ of all 
hadron-hadron scattering increase with $s$ as $s$ becomes large. 
It is also in disagreement with the Pomeranchuk theorem~\cite{Pomeranchuk58}. 
A  way to solve this problem is to add a trajectory with $\alpha_{0,P} > 1$ which does not correspond to any known particle
spectrum. 
This trajectory was called Pomeron trajectory with the exchanged particles carrying the quantum number  of ``vacuum''.
Within QCD, it  is natural to identify the Pomeron-exchange with the gluon-exchange, as investigated in 
Refs.~\cite{Low75,Nussinov75,DL88}. 
Accordingly, the the form factors $\beta^{13}_P(t)$ and $\beta^{24}_P(t)$ in Eq.~(\ref{eq:regge-f}) for the Pomeron trajectory
can be interpreted as the form factors due to the Pomeron-quark coupling.

An important step was taken by Donnachie and Landshoff~\cite{DL84} who demonstrated that very extensive hadron-hadron 
scattering data at high $s$ can be described very well by including only the Pomeron trajectory and assuming
\begin{enumerate}
\item  Pomeron-photon analogy~\cite{LP71a,JL74}: Pomeron couples with a quark in hadron like a photon and 
carries $s_P=+$ signature in Eq.~(\ref{eq:regge-f0}) to give it even $C$-parity.
\item The factorization approximation can be used to write the Pomeron-hadron form factor as
\begin{eqnarray}
\beta^{hh}_{P}(t)&=&\sum_{q}\beta_q F_h(t) ,
\end{eqnarray}
where $F_h(t)$ is the electromagnetic form factor of the hadron $h$, and $\beta_q$ is the Pomeron-quark coupling constant.
For the nucleon $h=N$, $F_N(t)=F_1(t)$ is the well determined Dirac form factor.
\end{enumerate}

With the above assumptions, Eq.~(\ref{eq:regge-f0}) for $A+B\rightarrow A+B$  scattering with even parity exchange
$n_s=+1$ at high $s$ then becomes
\begin{eqnarray}
T(s,t) &=&
\left[ \beta_{q_A} n_A F_A(t) \right] \left[ \beta_{q_B} n_B F_B(t) \right]
\frac{1+e^{-i\pi\alpha_{P}(t)}}{2\sin(\pi \alpha_P(t))}
\nonumber \\ && \mbox{} \times
(\alpha_{1,P}\,s)^{\alpha_P(t)}
\nonumber \\
&=&
\left[ \beta_{q_A} n_A F_A(t) \right]
\left[ \beta_{q_B} n_B F_B(t) \right] 
\frac{e^{-i\frac{\pi}{2}(\alpha_P(t)-1)}}{2\sin(\frac{\pi}{2}\alpha_P(t))}
\nonumber \\ && \mbox{} \times
(\alpha_{1,P}\,s)^{\alpha_P(t)} ,
\label{eq:regge-ff}
\end{eqnarray}
where $n_h$ is the number of quarks in the hadron $h$; $n_N=3$ for the nucleon, $n_\pi=2$ for the pion etc., 
$\beta_{q_A}$ is the quark-Pomeron coupling constant in hadron $A$;
$\beta_{q_A}=\beta_{q_B}=\beta_{u}=\beta_{d}$ for $\pi N$ and $pp$ scattering.

Taking $F_1(t)$ and $F_\pi(t)$ extracted from the experimental data, Donnachie and Landshoff~\cite{DL84} demonstrated that the total  
cross sections and the differential cross section data of $pp$, $p\bar{p}$ and $\pi^+p$ at small $|t|$ and high $s$ can be described within 
$10\%$ with the following parameters:
\begin{eqnarray}
\alpha_P(t)&=&1.08+0.25 t  \label{pom-traj} , \\
\beta_u^2&=&\beta_d^2=3.21~ \mbox{GeV}^{-2} ,
\end{eqnarray}
where $t$ is in the units of GeV$^2$.
The Pomeron-photon analogy had also been applied to get reasonably good description of diffractive dissociation reaction
$e+A \rightarrow e  + X$ with the assumption that the exchanged photon absorbed by one of the nucleons which is then elastically 
scattered from one of the other nucleon in the breakup system.  
The details can be found in Ref.~\cite{DDLN}.
We now turn to explaining how the Pomeron-exchange model of Donnachie and Landshoff can be applied to describe photoproduction 
of vector mesons.

\subsection{Photo-production of vector mesons}

To use the Pomeron-exchange amplitude defined by Eq.~(\ref{eq:regge-ff}) to describe the photo-production of vector mesons, 
one assumes that the incoming photon is converted into a vector meson $V$ by using the Vector Meson Dominance (VMD) 
model~\cite{Sakurai60,GZ61,Sakurai69,SS72a} defined by
\begin{eqnarray}
L_{\rm VMD}= \frac{em^2_V}{f_V}\phi_V^\mu(x)A_\mu(x) ,
\label{eq:vmd}
\end{eqnarray}
where $\phi_V(x)$ and $A_\mu(x)$ are the field operators for the vector meson $V$ and the photon, respectively. 
The coupling constant $f_V$ is traditionally determined by using Eq.~(\ref{eq:vmd}) to calculate the $V \to \gamma \to  e^+e^-$ 
decay width:
\begin{eqnarray}
\Gamma_{V\rightarrow e^+e^-}= \frac{1}{3}\alpha_{em}^2m_V\frac{4\pi}{f^2_V} .
\label{eq:vee-width}
\end{eqnarray}

With the Lagrangian of Eq.~(\ref{eq:vmd}), one can extend the $V+N \rightarrow V+N$ amplitude, defined by Eq.~(\ref{eq:regge-ff}), 
to describe $\gamma+ N \rightarrow V+N$. 
The resulting $Pom$-DL model is illustrated in the lower part of Fig.~\ref{fig:fig-pom-dl}.
The $Pom$-DL model had been applied in Refs.~\cite{WL12,WLZ19,Lee20,KLNO21} to fit the total cross section data of 
$\gamma + p \rightarrow V+p$ with $V=\rho,\phi,J/\Psi,\Upsilon$. 
The formulas used in Refs.~\cite{WL12,Lee20}, as given below, are used in the calculations presented in this paper.

We use the convention~\cite{GW} that the plane-wave state, $\ket{\mathbf{k}}$,  is normalized as  
$\braket{\mathbf{k}|\mathbf{k}'} = \delta(\mathbf{k}-\mathbf{k}' )$ and the $S$-matrix is related to the scattering $T$-matrix by
\begin{eqnarray}
S_{fi} =\delta_{fi}- 2\pi\, i\,\delta(E_i-E_f)\, T_{fi} .
\label{eq:s-matrix}
\end{eqnarray}
In the center of mass (CM) frame, the differential cross section of vector meson ($V$) photo-production
reaction, $\gamma (\mathbf{q})+N(-\mathbf{q})\rightarrow V(\mathbf{k})+N(-\mathbf{k})$, is calculated from
\begin{eqnarray}
\frac{d\sigma_{VN,\gamma N}}{d\Omega}(W)
&=& \frac{(2\pi)^4}{q^2}\rho_{VN}(k)\rho_{\gamma N}(q)
\nonumber \\ &&\hskip -1cm \mbox{} \times 
\frac{1}{4} \sum_{\lambda_V,m'_s}\sum_{\lambda_\gamma,m_s}
\left| \braket{\mathbf{k},\lambda_V m'_s|T_{VN,\gamma N}(W)|\mathbf{q},\lambda_\gamma  m_s} \right|^2 , 
\nonumber \\ &&
\label{eq:crst-gnjn}
\end{eqnarray}
where $m_s$ denotes the $z$-component of the nucleon spin, and $\lambda_V$ and $\lambda_\gamma$
are the helicities of vector meson $V$ and photon $\gamma$, respectively, and
\begin{eqnarray}
\rho_{VN}(k)&=&\frac{k\, E_V(k)E_N(k)}{W} \,,\\
 \rho_{\gamma N}(q)&=&\frac{q^2E_N(q)}{W}\,.
\label{eq:rhog}
\end{eqnarray}
The magnitudes of $k=|\mathbf{k}|$ and $q=|\mathbf{q}|$ are defined by the invariant mass
$W=q+E_N(q)=E_V(k)+E_N(k)$.

Within the $Pom$-DL model, the $\gamma +N \rightarrow V+N$ amplitude in Eq.~(\ref{eq:crst-gnjn}) is
\begin{eqnarray}
\braket{ \mathbf{k},\lambda_V m'_s | T_{VN,\gamma N}(W) | \mathbf{q},\lambda_\gamma m_s}
&=&
\braket{ \mathbf{k},\lambda_V m'_s | t^{\rm Pom}(W) | \mathbf{q},\lambda_\gamma m_s} ,
\nonumber \\
\label{eq:t-pom}
\end{eqnarray}
where
\begin{eqnarray}
&& \braket{ \mathbf{k},\lambda_V m'_s|t^{pom}(W)|\mathbf{q},\lambda_\gamma m_s} =
\nonumber \\
&&
\frac{1}{(2\pi)^3}\sqrt{\frac{ m_Nm_N }{4 E_{V}(\mathbf{k}) E_N(\mathbf{p}_f)
|\mathbf{q}|E_N(\mathbf{p}_i) }}\nonumber \\
&& \mbox{} \times 
[\bar{u}(p_f,m'_s)
\epsilon^*_\mu(k,\lambda_{V})\mathcal{M}^{\mu\nu}_\mathbb{P}(k,p_f,q,p_i)
\epsilon_\nu(q,\lambda_\gamma) u(p_i,m_s)] .
\nonumber\\
\label{eq:pomt}
\end{eqnarray}
In the above equation, we have defined four-momenta as $k=(E_V(\mathbf{k}),\mathbf{k})$, $p_f=(E_N(\mathbf{k}), -\mathbf{k})$,
$q_i=(q, \mathbf{q})$, $p_i=(E_N(\mathbf{q}), -\mathbf{q})$, and $\epsilon_\nu(q,\lambda_\gamma)$ is the polarization vector of 
photon, $\epsilon_\nu(k,\lambda_V)$ the polarization vector of vector meson $V$, and $u(p,m_s)$ is the nucleon spinor 
with the normalization $\bar{u}(p,m_s){u}(p,m'_s) = \delta_{m_s,m'_s}$.
The Pomeron-exchange amplitude $\mathcal{M}^{\mu\nu}_\mathbb{P}(k,p_f,q,p_i)$ can be written as
\begin{equation}
\mathcal{M}^{\mu\nu}_\mathbb{P}(k,p_f,q,p_i) = G_\mathbb{P}(s,t)
\mathcal{T}^{\mu\nu}_\mathbb{P}(k,p_f,q,p_i)
\label{eq:MP}
\end{equation}
with
\begin{eqnarray}
\mathcal{T}^{\mu\nu}_\mathbb{P}(q,p,q',p') &=& i \, 2 \frac{e \, m_V^2}{f_V}
 [2\beta_{q_{V}}F_V(t)][3\beta_{u/d} F_1(t)] 
 \nonumber \\ && \mbox{} \times
 \{ \slashed{q} g^{\mu\nu} - q^\mu \gamma^\nu \}  \, ,
\label{eq:pom-a}
\end{eqnarray}
where $M_V$ is the mass for the vector meson, and 
$f_V= 5.3$, $15.2$, $13.4$, $11.2$, $40.53$ for  $V=\rho, \omega, \phi, J/\Psi, \Upsilon$.
The parameters $\beta_{q_{V}}$ ($\beta_{u/d}$) define the coupling of the Pomeron with the quark $q_{V}$ ($u$ or $d$) in the vector meson 
$V$ (nucleon $N$).
In Eq.~(\ref{eq:pom-a}) a form factor for the Pomeron-vector meson vertex is also introduced with
\begin{eqnarray}
F_V(t)=\frac{1}{m_V^2-t} \left( \frac{2\mu_0^2}{2\mu_0^2 + m_V^2 - t} \right) ,
\label{eq:f1v}
\end{eqnarray}
where $t=(q-k)^2=(p_f-p_i)^2$. 
By using the Pomeron-photon analogy, the form factor for the Pomeron-nucleon vertex is defined by the isoscalar electromagnetic form 
factor of the nucleon as
\begin{equation}
F_1(t) = \frac{4m_N^2 - 2.8 t}{(4m_N^2 - t)(1-t/0.71)^2}.
\label{eq:f1}
\end{equation}
Here $t$ is in the unit of GeV$^2$, and $m_N$ is the proton mass.

The crucial ingredient of Regge phenomenology is the propagator $G_\mathbb{P}$ of the Pomeron in Eq.~(\ref{eq:MP}). 
Following Eq.~(\ref{eq:regge-ff}) and using $\sin[\frac{\pi}{2}\alpha_P(t)] \sim 1$ for $\alpha_P(t)\sim 1$ in the small $|t|$ region,
it is of the following form:
\begin{equation}
G_\mathbb{P} = \left(\frac{s}{s_0}\right)^{\alpha_P(t)-1}
\exp\left\{ - \frac{i\pi}{2} \left[ \alpha_P(t)-1 \right]
\right\} \,,
\label{eq:regge-g}
\end{equation}
where $s=(q+p_i)^2=W^2$, $ \alpha_P (t) = \alpha_0 + \alpha'_P t$, and $s_0=1/\alpha'_P$.

The amplitude of the $Pom$-DL model, as defined above, has the parameters: the quark-Pomeron coupling constants
$\beta_q$ for $q=u/d,s,c,b$, $\mu_0$ for the form factor $F_V(t)$ of the vector meson $V$, and $\alpha_0$ and $\alpha'_P$ 
for the Regge trajectory $\alpha_P(t)$.
The nucleon form factor $F_1(t)$ is given  in Eq.~(\ref{eq:f1}) and we set $s_0=1/\alpha'_P=0.25$ GeV from Donnachie and Landshoff. 
By fitting the data of $\rho^0$, $\omega$, $\phi$ photo-production~\cite{OL02}, we have determined:
$\mu_0=  1.1$ GeV$^2$, $\beta_{u/d}=2.07$ GeV$^{-1}$, $\beta_{s}=1.38$ GeV$^{-1}$, $\alpha_0=1.08$ for 
$\rho$ and $\omega$, $\alpha_0=1.12$ for $\phi$. 
For the heavy quark systems, we find that with the same $\mu_0^2$, $\beta_{u/d}$, and $\alpha'_P$, the $J/\Psi$ and $\Upsilon$ 
photo-production  data can be fitted by setting $\beta_c = 0.32$ GeV$^{-1}$ and $\beta_b = 0.45$ GeV$^{-1}$, and choosing a larger 
$\alpha_0=1.25$.

\begin{figure}[t] 
\centering
\includegraphics[width=0.95\columnwidth,angle=0]{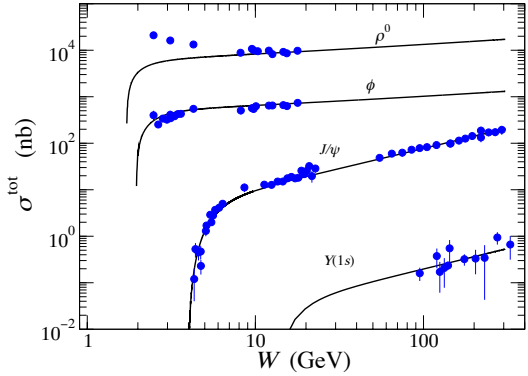}
\caption{Fits to the data of the total cross sections ($\sigma^{\rm tot}$) of photo-production of $\rho^0$, $\phi$, $J/\Psi$ and 
$\Upsilon(1s)$ on the proton target.
The solid curves are calculated from using the $Pom$-DL model.
Data are from Refs.~\cite{ZEUS-95b,H1-96,GHLL75,CLPS75,ZEUS-98,H1-00b,ZEUS-09,LHCb-12,ZEUS-95,SWGL82,BCEK73,AHHM76,EDLM79,BCEP82b,ZEUS-96b,BBHM78}.}
\label{fig:totcrst-all-v}
\end{figure}

In Fig.~\ref{fig:totcrst-all-v}, we see that the data for the $\phi$, $J/\Psi$, and $\Upsilon$ production can be described very well 
by the $Pom$-DL model.
On the other hand, the $\rho$ photo-production data at low energies clearly need other mechanisms such as the meson-exchange mechanisms 
illustrated in Refs.~\cite{PL96,PL97} or the mechanisms due to the $a_2$ and $f_2$ Regge trajectories as included in Refs.~\cite{DL95,OL04}.
It appears that the slop parameter $\alpha_0$ for the energy-dependence of the diffractive production of heavy quarks ($c$ and $b$) is rather 
different from that for light quarks ($u$, $d$, $s$).
This was interpreted~\cite{DL98} as the presence of second Pomeron.  
As with much of Pomeron phenomenology, it will be good to understand this observation within QCD.

\begin{figure}[t]
\centering
\includegraphics[width=0.9\columnwidth,angle=0]{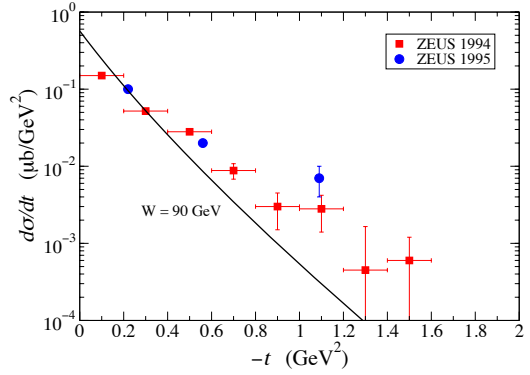}
\includegraphics[width=0.9\columnwidth,angle=0]{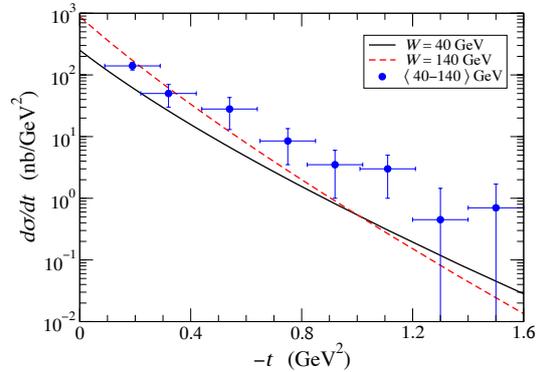}
\caption{Differential cross sections from the $Pom$-DL model are compared with ZEUS data~\cite{ZEUS-95b,H1-96,GHLL75,CLPS75}.
Upper: at $W= 90$~GeV; Lower: at $W= 40$~GeV (solid curve) and $140$ GeV (dashed curve), data are from averaging the data in the
range of $W=40$-140 GeV .}
\label{fig:dsdt-w90-pom}
\end{figure}

\begin{figure}[t]
\centering
\includegraphics[width=0.9\columnwidth,angle=0]{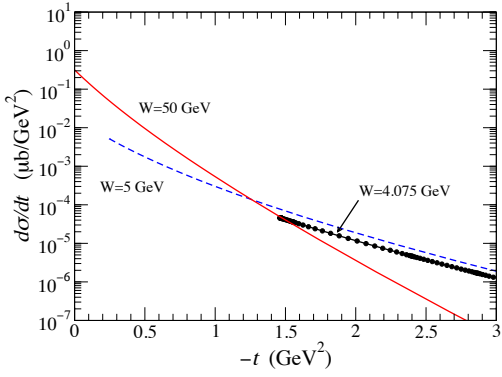}
\caption{Differential cross sections at  $W=4.075$, $5$, $50$ GeV calculated from the $Pom$-DL model .}
\label{fig:dsdt-allw-pom}
\end{figure}

In Fig.~\ref{fig:dsdt-w90-pom}, we show that the predicted differential cross sections at $W=40$-$140$~GeV are 
consistent with the data (averaged over the cross sections in a range of $W$), while some refinements of the model are needed to fit data at large $-t$.
It is worthwhile to use the $Pom$-DL model to illustrate that in the near threshold energy region $W \lesssim 5$~GeV, the cross sections are from  the
large momentum-transfer region where the Regge phenomenology may not be applicable. 
This can be seen in Fig.~\ref{fig:dsdt-allw-pom}. 
At the very near threshold $W=4.075$~GeV, the momentum transfers are in the $-t > 1.4  $~GeV$^2$ region.
We thus expect that the $Pom$-DL model needs to be improved to fit the JLab data. 
This can be seen in Fig.~\ref{fig:pom-dl} where the predicted  total cross sections are shown to be much smaller than the JLab data.

\section{$Pom$-pot model}
\label{sec:Pom+pot}

\begin{figure}[t] 
\centering
\includegraphics[width=0.9\columnwidth,angle=0]{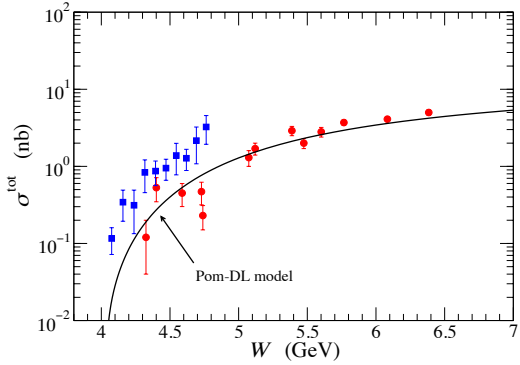}
\caption{Total cross sections calculated from  the $Pom$-DL model are compared with the data. 
Solid squares are the JLab  data~\cite{GlueX-19}.}
\label{fig:pom-dl}
\end{figure}

The results  shown in Fig.~\ref{fig:pom-dl} is not unexpected since the use of Pomeron-photon analogy only accounts for parts of gluon-exchange 
mechanisms which may not be the dominant mechanism in the near threshold region. 
Adding the non-perturbative multi-gluon exchange mechanisms as included in the $J/\Psi$-N potentials $v_{J/\Psi N,J/\Psi  N}$ extracted from LQCD 
may solve the problem.
By also using the VMD assumption, the amplitude ($t^{\rm pot}$) due to a potential $v_{J/\Psi N,J/\Psi  N}$ is illustrated in Fig.~\ref{fig:fig-pom-pot}.

To proceed, we note that the VMD coupling constant $1/f_V$ is determined by the $J/\Psi \rightarrow \gamma \rightarrow  e^+e^-$ decay at 
$q^2=m^2_{J/\Psi}\sim 9$~GeV$^2$ which is far from $q^2=0$ for the photo-production process.  
It is therefore necessary to include an offf-shell factor $F_{\rm off}(q^2)$ with $F_{\rm off}(m^2_V)=1$ to correct this. 
This phenomenological procedure is needed since VMD for $J/\Psi$ is questionable, as recently claimed in Ref.~\cite{YBCR22}.  
The amplitude from extending the $Pom$-DL model to include the amplitude $t^{\rm pot}$ shown in Fig.~\ref{fig:fig-pom-pot} due to 
$v_{J/\Psi N,J/\Psi  N}$ has the  following form:
\begin{eqnarray}
\braket{\mathbf{k},\lambda_V m'_s|T_{VN,\gamma N}(W)|\mathbf{q},\lambda_\gamma m_s}
&=&
\nonumber \\ && \hskip -1.5cm
 \braket{\mathbf{k},\lambda_V m'_s|t^{\rm Pom+pot}(W)|\mathbf{q},\lambda_\gamma m_s}
 \label{eq:pompot-t}
\end{eqnarray}
with
\begin{eqnarray}
\braket{ \mathbf{k},\lambda_V m'_s | t^{\rm Pom+pot}(W) | \mathbf{q},\lambda_\gamma m_s}
&=& \braket{ \mathbf{k},\lambda_V m'_s|t^{\rm Pom}(W)|\mathbf{q},\lambda_\gamma m_s}
\nonumber \\  &&  \mbox{} \hskip -1cm
+ \braket{\mathbf{k},\lambda_V m'_s|t^{\rm pot}(W)|\mathbf{q},\lambda_\gamma m_s} , 
\end{eqnarray}
where $\braket{ \mathbf{k},\lambda_V  m'_s | t^{\rm Pom}(W) | \mathbf{q},\lambda_\gamma m_s}$ has been given in Eq.~(\ref{eq:pomt}), 
and we follow the dynamical formulation of Refs.~\cite{SL96,MSL06,JLMS07,KNLS13} to write
\begin{eqnarray}
\braket{ \mathbf{k},\lambda_V m'_s|t^{\rm pot}(W)|\mathbf{q},\lambda_\gamma m_s}
&=&
\braket{ \mathbf{k},\lambda_V m'_s|t_{VN,VN}(W)|\mathbf{q},\lambda_\gamma m_s}
\nonumber \\ && \mbox{} \hskip -2.2cm   \times
\frac{1}{W-E_N(q)-E_V(q)+i\epsilon} 
\nonumber \\ && \hskip -2.2cm \times
\left[ F_{\rm off}(0)\frac{em^2_V}{f_V}\frac{1}{(2\pi)^{3/2}}\frac{1}{\sqrt{2q}}\frac{1}{\sqrt{2E_V(q)}} \right] .
\label{eq:lqcd-t-1}
\end{eqnarray}
Here, the $J/\Psi$-N scattering amplitude $t_{VN,V N}(W)$ is calculated from potential $v_{VN,VN}$ by using the 
following Lippmann-Schwinger equation,
\begin{eqnarray}
t_{VN,V N}(W) = v_{VN,VN} + v_{VN,VN}\frac{1}{W-H_0+i\epsilon}t_{VN,V N}(W) .
\label{eq:lpeq}
\end{eqnarray}
The above equation is solved in partial-wave representation.
For a central potential, such as $v_{J/\Psi N, J/\Psi N}=v_0\frac{-\alpha r}{r}$ extracted from LQCD~\cite{KS10b}, we have
\begin{eqnarray}
\braket{ \mathbf{k},m_Vm_s|t_{VN,VN}(W)|\mathbf{k}^{\,\,'},m'_V m'_s} &=&
\nonumber \\ && \mbox{} \hskip -2.7cm 
\sum_{LM}Y^*_{LM}(\hat{k})t_L(k,k')Y_{LM}(\hat{k}')\delta_{m_V,m'_V}\delta_{m's_,m_s} .
\end{eqnarray}
The matrix element $t_L(k,k',W)$  is then from solving
\begin{eqnarray}
t_L(k,k',W) &=& v_L(k,k')
\nonumber \\ && \mbox{}
+ \int q^2\,dq\,v_L(k,q)\frac{1}{W-E_N(q)-E_V(q)+i\epsilon}
\nonumber \\ && \mbox{} \qquad \times
t_L(q,k',W) , 
\end{eqnarray}
where
\begin{eqnarray}
v_L(k,k')=\frac{2}{\pi}\int r^2\,dr\,  j_L(kr) v_{V N, V N}(r) j_L(k'r) .
\end{eqnarray}

\begin{figure}[t] 
\centering
\includegraphics[clip,width=0.9\columnwidth,angle=0]{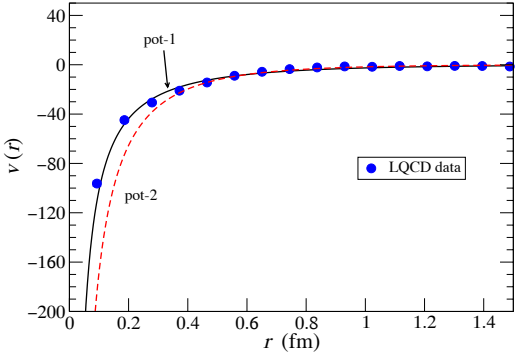}
\caption{The $J/\Psi$-N potential $v_{J/\Psi N, J/\Psi N}(r)=v_0\frac{e^{-\alpha r}}{r}$ extracted from the LQCD calculations of 
Refs.~\cite{KS10b,sasaki-1}.}
\label{fig:v-sasaki-lqcd}
\end{figure}

\begin{figure}[t]
\centering
\includegraphics[width=0.9\columnwidth,angle=0]{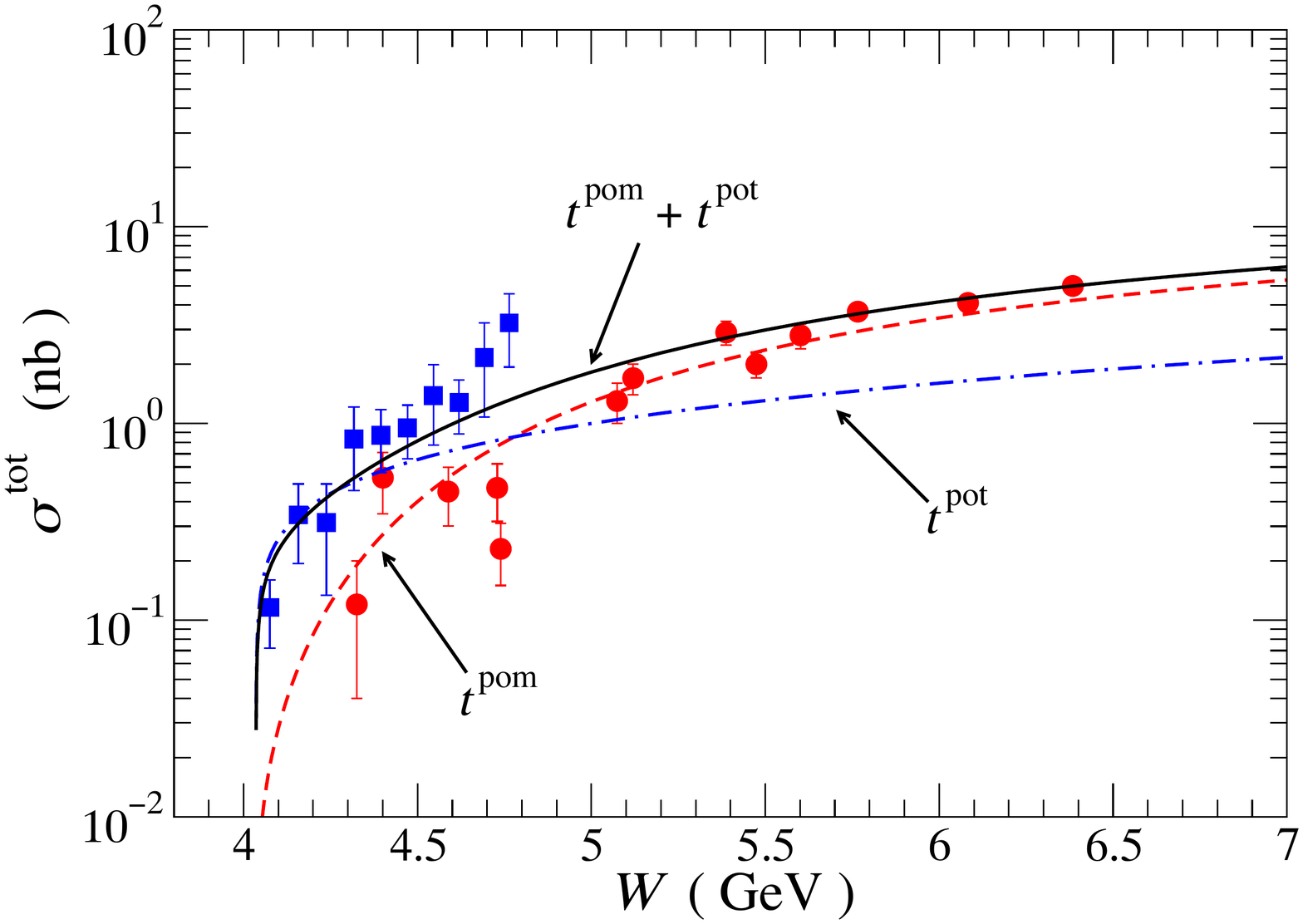}
\caption{The total cross sections of $\gamma+p\rightarrow J/\Psi+p$. 
$t^{\rm Pom}$ ($t^{\rm pot}$) indicates the cross sections calculated from keeping only $t^{\rm Pom}$ ($t^{\rm pot}$) term in Eq.~(\ref{eq:pompot-t}). 
$t^{\rm Pom}+t^{\rm pot}$ indicate the cross sections calculated from the total amplitude. 
}
\label{fig:p-totcrst-pm}
\end{figure}

The  $Pom$-pot model has been applied in Ref.~\cite{Lee20} by using $v_{J/\Psi N, J/\Psi N}$ extracted from the LQCD calculations of 
Refs.~\cite{KS10b,sasaki-1}.
As shown in Fig.~\ref{fig:v-sasaki-lqcd}, the LQCD data (solid circles) can be fitted by two sets of parameters: 
(1) $v_0=-0.06$, $\alpha=0.3$~GeV (pot-1), (2) $v_0=-0.11$, $\alpha=0.5$~GeV (pot-2).
It is found that the JLab data can be fitted by choosing the off-shell factor $F_{\rm off}(0)=0.7$ $(0.4)$ for pot-1 (pot-2).
The results from using pot-1 are shown in Fig.~\ref{fig:p-totcrst-pm}.
We see that the amplitude $t^{\rm pot}$ from $v_{J/\Psi N, J/\Psi N}$ interferes with $t^{\rm Pom}$ from the $Pom$-DL model
to fit the JLab data in $W < 5$~GeV region, while the fits at higher energies are also improved.

\begin{figure}[t]
\centering
\includegraphics[width=0.9\columnwidth,angle=0]{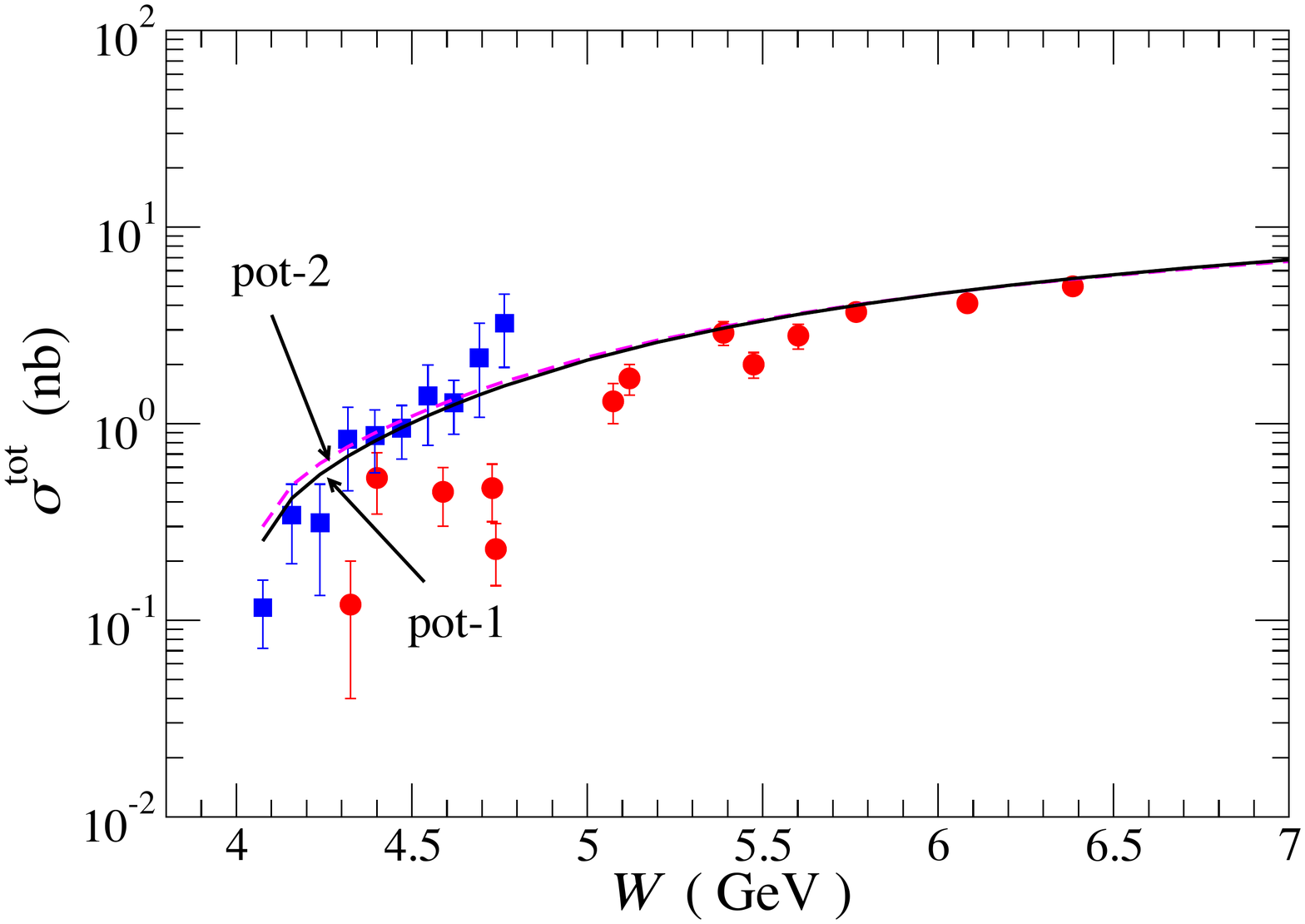}
\includegraphics[width=0.9\columnwidth,angle=0]{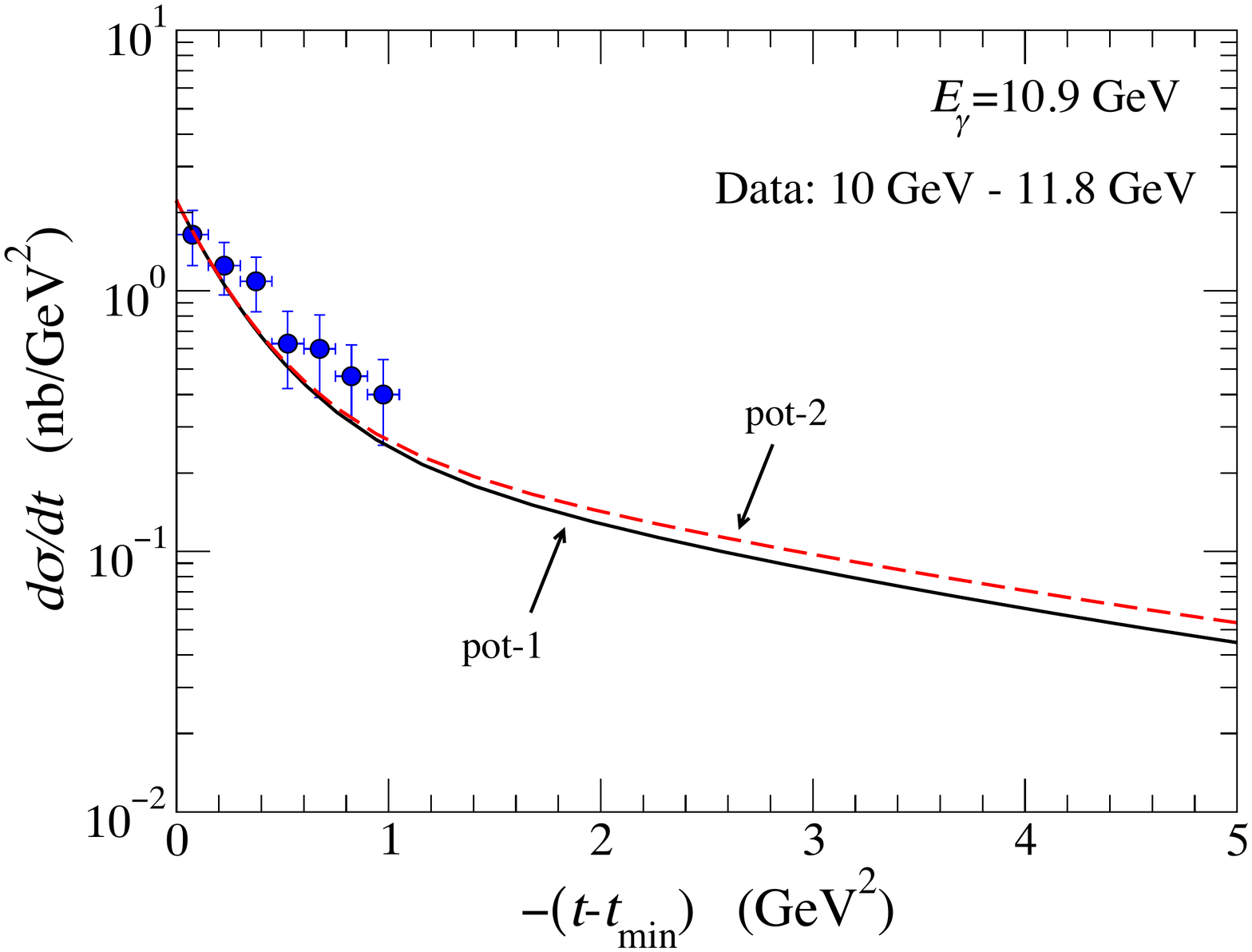}
\caption{ 
Cross sections of $\gamma+p\rightarrow J/\Psi+p$ from using LQCD potentials pot-1 and pot-2.
Upper: total cross sections; Lower: differential cross sections.
}
\label{fig:p-totcrst-pm-1}
\end{figure}

In the upper part of Fig.~\ref{fig:p-totcrst-pm-1}, we see that the fit to the JLab total cross section data by using pot-2 is similar to that from pot-1. 
In the lower part, we see that the both models give equally good descriptions of the differential cross sections.
The $Pom$-pot model can describe the data from threshold to very high energies of $W=300$~GeV.
Hence it can be used to predict the cross sections of $J/\Psi$ photo-production from nuclei for investigating the  production  of nuclei with hidden
charms~\cite{BD88a,BSD90,GLM00,BSFS06,WL12}, the gluon distributions in nuclei, and the nucleon-nucleon short-range correlations  
as discussed in Ref.~\cite{HSXY19}.

\section{Models of PQCD}
\label{sec:pqcd}

In this section, we give the formulas for using $GPD$-based (Fig.~\ref{fig:gpd}) and $2g+3g$ (Fig.~\ref{fig:2g3g}) models 
within the formulation defined by Eqs.~(\ref{eq:s-matrix})-(\ref{eq:rhog}). 

\subsection { $2g+3g$ model}

The amplitudes of the 
$2g+3g$ model, as defined by Eqs.~(\ref{eq:2g})-(\ref{eq:3g}), can be written as~\cite{WL12}
\begin{eqnarray}
\braket{ \mathbf{k},\lambda_{J/\Psi}m'_s|T_{J/\Psi\,N,\gamma N}(W)|\mathbf{q},\lambda_\gamma m_s} =
&&
\nonumber \\ && \mbox{} \hskip -2.5cm
\braket{ \mathbf{k},\lambda_{J/\Psi}m'_s|t^{2g+3g}(W)|\mathbf{q},\lambda_\gamma m_s} ,
\label{eq:2g3gt-0}
\end{eqnarray}
where by defining the four-momenta as $k=(E_{J/\Psi}(\mathbf{k}),\mathbf{k})$, $p_f=(E_N(\mathbf{k}), -\mathbf{k})$,
$q_i=(q, \mathbf{q})$, $p_i=(E_N(\mathbf{q}), -\mathbf{q})$, the amplitude can be written as 
\begin{eqnarray}
&& \braket{ \mathbf{k},\lambda_{J/\Psi} m'_s|t^{2g+3g}(W)|\mathbf{q},\lambda_\gamma m_s}
\nonumber \\ & =&
\frac{1}{(2\pi)^3}\sqrt{\frac{ m_Nm_N }{4 E_{J/\Psi}(\mathbf{k}) E_N(\mathbf{p}_f) |\mathbf{q}|E_N(\mathbf{p}_i) }}
%\nonumber  \\ && \mbox{} \times 
\frac{4\sqrt{\pi}}{\sqrt{6}} \frac{qW}{m_N}[M_{2g}+M_{3g}]
\nonumber \\
\label{eq:2g3gt}
\end{eqnarray}
with
\begin{eqnarray}
M_{2g}&=&\frac{A_{2g}}{4\sqrt{\pi}}\frac{1-x}{R\,m_{J/\Psi}}e^{bt/2} , \\
M_{2g}&=&\frac{A_{3g}}{4\sqrt{\pi}}\frac{1}{R^2\,m^2_{J/\Psi}}e^{bt/2} ,\\
x&=&\frac{2m_Nm_{J/\Psi}+m^2_{J/\Psi}}{W^2-m_p^2} ,
\label{eq:2g3gt-1}
\end{eqnarray}
where $R=1$~fm, $b=1.13$~GeV$^{-2}$ are taken from Ref.~\cite{BCHL01}.
Substituting Eqs.~(\ref{eq:2g3gt-0})-(\ref{eq:2g3gt}) into Eq.~(\ref{eq:crst-gnjn}), we obtain 
\begin{eqnarray}
\frac{d\sigma_{J/\Psi\, N,\gamma N}}{d\Omega} = \frac{qk}{\pi}
\left[ \frac{d\sigma_{2g}}{dt} + \frac{d\sigma_{3g}}{dt} 
+ C\times[2 M_{2g}\,M_{3g}] \right],
\end{eqnarray}
where $\frac{d\sigma_{2g}}{dt}$ and $\frac{d\sigma_{3g}}{dt}$ are identical to those given in Eqs.~(\ref{eq:2g}) and (\ref{eq:3g}), and $C$ is a kinematic factor.
The term $C\times[2 M_{2g}\,M_{3g}]$ is from the interference between the amplitudes $M_{2g}$ of $2g$-exchange and $M_{3g}$ of $3g$-exchange.
This term makes our approach slightly different from Ref.~\cite{BCHL01}.
This modification is necessary for our later study of $N^*$ in which the $N^*$ interference with the non-resonant amplitudes is crucial.

 \begin{figure}[t]
\centering
\includegraphics[width=0.9\columnwidth,angle=0]{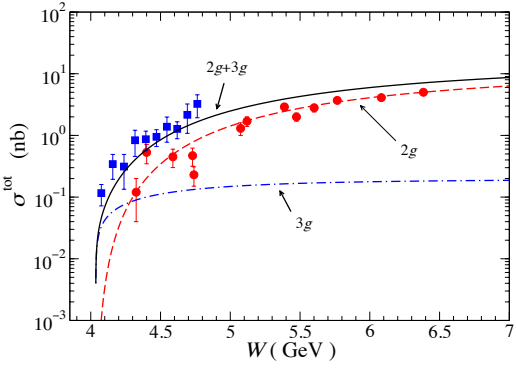}
\caption{Total cross sections of $\gamma +p\rightarrow J/\Psi+p$ calculated from $2g+3g$ model. 
$2g$ ($3g$) is the contribution from two-gluon  (three-gluon) exchange amplitudes of Eq.~(\ref{eq:2g3gt}).
}
\label{fig:p-totcrst-2g3g}
\end{figure}

We follow Ref.~\cite{BCHL01} to determine the parameters $A_{2g}$ and $A_{3g}$ by fitting the total cross section data.
As shown in Fig.~\ref{fig:p-totcrst-2g3g}, we find the old data (solid circles) can be fitted by the two-gluon exchange ($2g$) model (dashed  curve)
with $(A_{2g}=0.023$~MeV$^{-2}$, $ A_{3g}=0)$.
The JLab data can be fitted (solid curve) by adding the $3g$ exchange contribution (dash-dot curve) with 
$(A_{2g}=0.028$~MeV$^{-2}$ and 
$A_{3g}=2000$~MeV$^{-2}$).

\begin{figure}[t]
\centering
\includegraphics[width=0.9\columnwidth,angle=0]{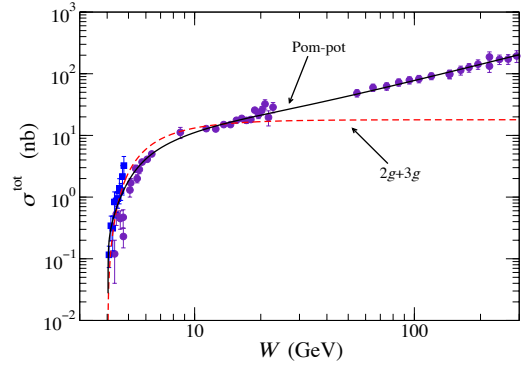}
\includegraphics[width=0.9\columnwidth,angle=0]{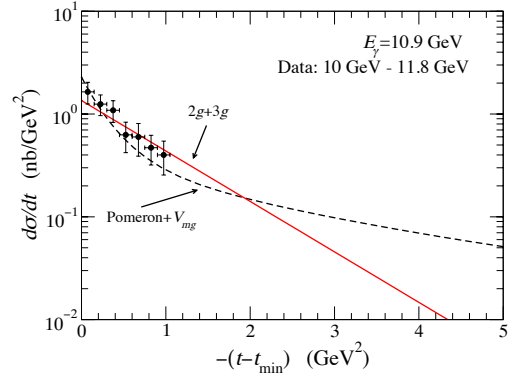}
\caption{Comparisons of the cross sections from $Pom$-DL and $2g+3g$ models.
Upper: total  cross sections , Lower: differential cross sections.
}
\label{fig:dsdt-comp-pomlqcd-2g3g-1}
\end{figure}
%FIGURE<<<

The main difference between the $Pomi$-pot model and the $2g+3g$ model is at high energies, as shown in the upper part of Fig.~\ref{fig:dsdt-comp-pomlqcd-2g3g-1}.
It remains to be explored whether the $2g+3g$ model can be improved to also fit the data at high energies.
Their difference can also be distinguished in differential cross sections at large $t$, as illustrated in the lower part of Fig.~\ref{fig:dsdt-comp-pomlqcd-2g3g-1}.
Clearly more precise data are needed to test the predictions.

\subsection{$GPD$-based model}

In this subsection, we explain the derivations of the cross section in Eq.~(\ref{eq:crst-gpd}) from Eq.~(\ref{eq:gpd-exact}) within the $GPD$-based model. 
It is convenient to perform the calculation of the amplitude of $\gamma ({q})+p(P)\rightarrow J/\Psi (K) +  p(P')$ in the CM frame within which the initial nucleon 
momentum ${\bf P}$ is in the $z$-direction and the final nucleon in the $x$-$z$ plane: 
\begin{eqnarray}
P&=&(P^0,0,0,|{\bf P}|),\nonumber \\
q&=&(q^0,0,0,-|{\bf P}|),\nonumber \\
P'&=&(P^{'0}, |{\bf P}'|\sin\theta,0, |{\bf P}'|\cos\theta),  \nonumber \\
K&=&(K^0,-|{\bf P}'|\sin\theta,0, -|{\bf P}'|\cos\theta), \nonumber
\end{eqnarray}
where $\theta$ is the angle between ${\bf  P}'$ and $z$-axis.
The skewness $\xi$ is then defined by
\begin{eqnarray}
\xi&=&-\frac{\Delta\cdot n}{2\bar{P}\cdot n} ,
\end{eqnarray}
where  $\Delta=P'-P$, and the light-cone vector $n$ is defined as $n\equiv\frac{1}{\sqrt{2\bar{P}^+}}(1,{\bf 0}_\bot,-1)$ with the average nucleon momentum 
$\bar P\equiv (P+P')/2$ such that $n\cdot \bar P =1$.
We then have
\begin{eqnarray}
\xi &=& \frac{P^+-P^{'+}}{P^++P^{'+}}
\nonumber \\
&=& \frac{ [E_N({\bf  P}) + |{\bf P}|] - [E_N({\bf P}')+ |{\bf  P}'|\cos\theta]}
{ [ E_N({\bf  P}) + |{\bf P}| ]+ [E_N({\bf P}')+ |{\bf  P}'|\cos\theta ]} ,
\end{eqnarray}
where  $P^+=\frac{1}{\sqrt{2}}(P^0+P^3)$.

\begin{figure}[t]
\centering
\includegraphics[width=0.9\columnwidth,angle=0]{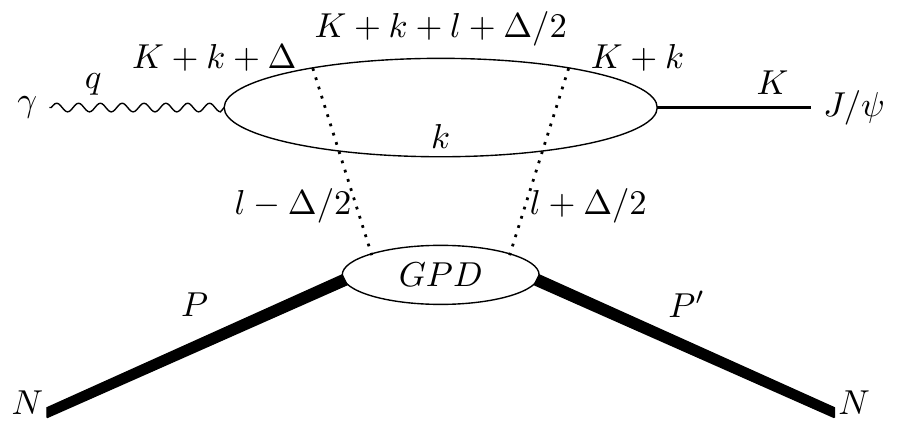}
\vskip 1cm
\includegraphics[width=0.9\columnwidth,angle=0]{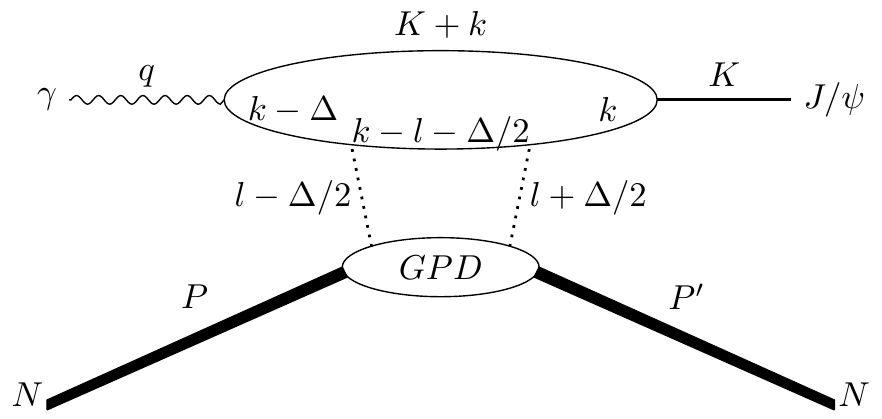}
\caption{ The momentum variables of the two-gluon exchange amplitudes $I^b_{\mu\nu}$ (upper) and $I^d_{\mu\nu}$ (lower) of Eq.~(\ref{eq:gpd-exact-d}).
}
\label{fig:gpd-d}
\end{figure}

To illustrate the derivations of the formulas in Ref.~\cite{GJL21}, it is sufficient to only consider two of the four possible two-gluon-exchange amplitudes, 
as illustrated in Fig.~\ref{fig:gpd-d}.
The  amplitude can then be written as 
\begin{eqnarray}
t_{J/\Psi\,N,\gamma N}(P' K; P q)
 &=&
\nonumber \\ && \mbox{} \hskip -2.5cm 
  (ig)^2  \int \frac{d^4k \, d^4 l}{(2\pi)^8} \textstyle
\braket{ P' | A^\mu(-l-\frac{\Delta}{2}) A^\nu(l-\frac{\Delta}{2}) | P }
\nonumber \\ && \mbox{} \hskip -2.5cm  \times 
\mbox{Tr} \left\{\left[I^b_{\mu\nu}(K,k,l,\Delta,\epsilon_\gamma)
+I^d_{\mu\nu}(K,k,l,\Delta,\epsilon_\gamma)\right]
\Psi(K,k)\right\} \,,\nonumber \\
&& \label{eq:gpd-exact-d}
\end{eqnarray}
where $\braket{P'|A^\mu(-l-\frac{\Delta}{2}) A^\nu(l-\frac{\Delta}{2})|P}$ describes the distribution of the two exchanged-gluons in the nucleon,  
$\epsilon_\gamma$ is the photon polarization vector,
\begin{eqnarray}
\Psi(K,k)=\left<\Psi_{K}\right|T\{\psi_c(k)\bar{\psi}_{\bar{c}}(K+k)\}\left|0\right>
\label{eq:bs-wf}
\end{eqnarray}
is the $c\bar{c}$ wavefunction of $J/\Psi$, and
\begin{eqnarray}
I^{\mu\nu}_b&=&
\gamma^\mu\frac{i}{\slashed{K}+\slashed{k}+\slashed{l}+\slashed{\Delta}/2-m_c+i\epsilon}
\gamma^\nu 
\nonumber\\ && \mbox{} \times
\frac{i}{\slashed{K}+\slashed{k}+\slashed{\Delta}-\slashed{\Delta}/2-m_c+i\epsilon}\slashed{\epsilon}_\gamma ,
\label{eq:gp-b} \\
I^{\mu\nu}_d&=&
\slashed{\epsilon}_\gamma\frac{i}{\slashed{k}-\slashed{\Delta}-m_c+i\epsilon}
\gamma^\nu \frac{i}{\slashed{k}-\slashed{l}-\slashed{\Delta}/2-m_c+i\epsilon}\gamma^\mu
\label{eq:gp-d}
\end{eqnarray}
are the  quark propagators in Fig.~\ref{fig:gpd-d}.

The  calculations of Eqs.~(\ref{eq:gpd-exact-d})-(\ref{eq:gp-d}) are simplified by using the following approximations :
\begin{enumerate}
\item Use  the threshold kinematics that the final $J/\Psi$ and nucleon are at rest to simplify the loop integration over 
quark propagators by setting 
\begin{eqnarray}
 \Delta&\sim&-(m_V,0,0,m_V/2) , \label{eq:appr-1} \\
l&\sim&x(m_V/2,0,0,m_V/2) , \label{eq:appr-2}
\end{eqnarray}
where  $-1 \leq x\leq  +1$ is the average momentum fraction of the two exchanged gluons.
\item Use a non-relativistic approximation to write
\begin{eqnarray}
\Psi(K,k) &\sim& \frac{u(k)\bar{v}(K+k)}{2E_c(k)}\tilde{\phi}(k)
\nonumber  \\
&\rightarrow& \frac{1}{4}(1-\slashed{\beta}_V)\frac{1}{\sqrt{2}}\epsilon_V
(1+\slashed{\beta}_V)
\nonumber \\ && \mbox{} \times
(2\pi)^4\delta^{(4)}(k+K/2)\phi_{NR}(0) ,
\label{eq:non-wf}
\end{eqnarray}
where $\beta_V \equiv K/m_V$ and the non-relativistic wavefunction at relative distance between two quarks $r=0$
is given by $\phi_{NR}(0)$.
\item Taking the light front gauge $A^+=0$, only $A^{\perp}$ of the gluon field contributes to the leading twist.
Therefore, only the terms with $\mu$, $\nu$ in the transverse direction are kept in the calculation of Eq.~(\ref{eq:gpd-exact-d}).
\end{enumerate}
By using Eqs.~(\ref{eq:appr-1})-(\ref{eq:non-wf}), Eq.~(\ref{eq:gpd-exact-d}) becomes
\begin{eqnarray}
t_{J/\Psi N,\gamma N}(P' K; P q)
 &=& (ig)^2\phi_{NR}(0)  
 \nonumber \\ && \hskip -2cm \mbox{} \times
 \int \frac{d^4 l}{(2\pi)^4} \textstyle
\braket{P' | A^\mu(-l-\frac{\Delta}{2}) A^\nu(l-\frac{\Delta}{2}) P }
\nonumber \\ && \hskip -1.5cm \mbox{} \times 
2\frac{1}{\sqrt{2}m^2_c} \left[ g^{\mu\nu}(\epsilon_\gamma\cdot\epsilon_V)
+\epsilon^\nu_\gamma\epsilon^\mu_V-\epsilon^\mu_\gamma\epsilon^\nu_V \right]_\perp ,
\label{eq:amp-final}
\end{eqnarray}
where the $\perp$ subscript indicates that only the  transverse components are non-zero.

By using the amplitude of Eq.~(\ref{eq:amp-final}), the differential cross section can be written as
\begin{eqnarray}
\frac{d\sigma}{dt} = \frac{\alpha_{EM}e^2_Q}{4(W^2-m^2_N)^2}\frac{(16\pi\alpha_S)^2}{3m^3_{J/\Psi}}
|\phi_{NR}(0)|^2|G(t,\xi)|^2\,,
\end{eqnarray}
where  
\begin{eqnarray}
G(t,\xi)&=&\frac{1}{2\xi}\int^{+1}_{-1}dx \left[\frac{1}{x+\xi-i\epsilon}-\frac{1}{x-\xi+i\epsilon}\right]
\nonumber \\ && \mbox{} \times
F_g(x,\xi,t) .
\label{eq:gtxi}
\end{eqnarray}
Here we have followed the standard definition of gluon GPD,
\begin{eqnarray}
F_g(x,\xi,t) &=& \frac{1}{(\bar{P}^+)^2} \int \frac{d\lambda}{2\pi} e^{i\lambda x}
\nonumber \\ && \mbox{} \times
\textstyle
\braket{P' | \mbox{Tr} \{F^{+i}(-\frac{\lambda n}{2})
F^{+}_i(-\frac{\lambda n}{2})\} | P },
\label{eq:gpd-def}
\end{eqnarray}
where we have defined in terms of gluon field $A^\mu$ of Eq.~(\ref{eq:gpd-exact-d})
\begin{eqnarray}
&& \frac{1}{(\bar{P}^+)^2} \textstyle [ F^{+i} (-\frac{\lambda n}{2}) F^{+}_i(-\frac{\lambda n}{2})] 
\nonumber \\ &=& \textstyle
(x+\xi)A^i(-\frac{\lambda n}{2})(x-\xi)A_i(-\frac{\lambda n}{2}) .
\end{eqnarray}

Expanding the  propagators in Eq.~(\ref{eq:gtxi}) as
\begin{eqnarray}
&& \frac{1}{2\xi}\left[\frac{1}{x+\xi-i\epsilon}
-\frac{1}{x-\xi+i\epsilon}\right]_{\epsilon\rightarrow 0}
\nonumber \\
&=& \frac{1}{2\xi}\frac{1}{\xi} \left[ \left(1+\frac{x}{\xi} \right)^{-1} + \left( 1-\frac{x}{\xi} \right)^{-1} \right]
\nonumber\\
&=& \frac{1}{2\xi}\frac{1}{\xi} \left[ \left( 1-\frac{x}{\xi}+\frac{x^2}{\xi^2}-\dots \right)
+ \left(1+\frac{x}{\xi}+\frac{x^2}{\xi^2}- \dots \right) \right]
\nonumber  \\
&=&\frac{1}{\xi^2} \left[1+ \left(\frac{x}{\xi} \right)^2 + \dots \right] ,
\end{eqnarray}
$G(t,\xi)$ is then expressed in terms of the even moments of the GPDs, 
\begin{eqnarray}
G(t,\xi)&=&\sum_{n=0}^{\infty}G^{(n)}(t,\xi) 
\end{eqnarray}
with
\begin{eqnarray}
G^{(n)}(t,\xi)=\frac{1}{\xi^{2n+2}}\int_{-1}^{+1} dx\,x^{2n}F_g(x,\xi,t)\,.
\end{eqnarray}
The  GPD defined by Eq.~(\ref{eq:gpd-def}) is parameterized as
\begin{eqnarray}
F_g(x,\xi,t)&=&\frac{1}{2\bar{P}^+}
\biggl[ H_g(x,\xi,t)\bar{u}(P')\gamma^+u(P)
\nonumber \\ && \mbox{}
+E_g(x,\xi,t)\bar{u}(P')\frac{i\sigma^{+\alpha}\Delta_\alpha}{2M_N}u(P) \biggr] .
\end{eqnarray}
Keeping only the $n=0$ moment, we then have
\begin{eqnarray}
|G(t,\xi)|^2 &\rightarrow& |G^{(0)}(t,\xi)|^2
\nonumber \\
&=&\frac{1}{\xi^4} \biggl[ \left(1-\frac{t}{4M^2_N} \right) E^2_2(t,\xi)
\nonumber \\ && \mbox{} \qquad 
-2E_2(x,\xi)(H_2(x,\xi)+E_2(x,\xi))
\nonumber \\ && \mbox{} \qquad
+(1-\xi^2)(H_2(x,\xi)+E_2(x,\xi))^2 \biggr] ,
\end{eqnarray}
where 
\begin{eqnarray}
H_2(x,\xi)&=&\int_0^1  dx\, H_g(x,\xi,t)= A_g(t) +(2\xi)^2 C_g(t) , \nonumber \\
E_2(x,\xi)&=&\int_0^1  dx\, E_g(x,\xi,t)=B_g(t)-(2\xi)^2 C_g(t) .
\end{eqnarray}

\begin{figure}[t]
\centering
\includegraphics[width=0.9\columnwidth,angle=0]{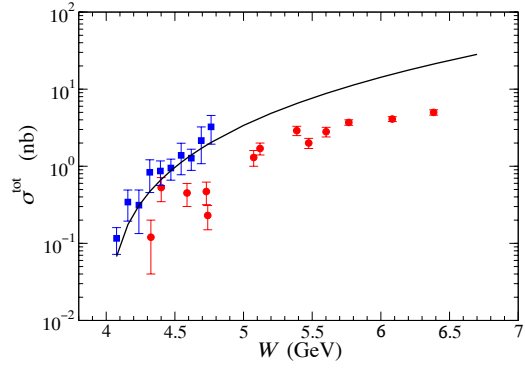}
\includegraphics[width=0.9\columnwidth,angle=0]{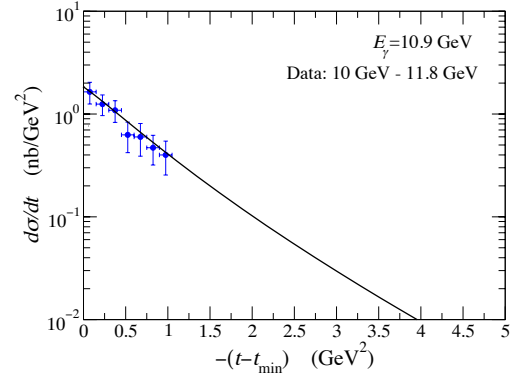}
\caption{$GPD$-based model. 
Upper: total cross sections, Lower: differential cross sections. }
\label{fig:gpd-based}
\end{figure}

To use the results from the LQCD calculation of Ref.~\cite{SD18}, one chooses the following dipole parameterization:
\begin{eqnarray}
A_g(t)&=&\frac{A_g(0)}{(1-\frac{t}{m^2_A})^2} , \\
C_g(t)&=&\frac{C_g(0)}{(1-\frac{t}{m^2_c})^2} , \\
B_g(t)&\sim& 0
\end{eqnarray}
with
\begin{eqnarray}
m_A&=& 1.13 \mbox{ GeV}, \qquad
m_c = 0.48 \mbox{ GeV} ,\\
A_g(0)&=&0.58, \qquad \qquad
C_g(0)= -1.0 .
\end{eqnarray}
To fit the JLab data, the above parameters were changed in Ref.~\cite{GJL21} as $m_A=1.13\rightarrow (1.64 \pm 0.11)$ 
and $C_q(0)=-1\rightarrow (-0.84 \pm 0.82)$.
The results are in Fig.~\ref{fig:gpd-based}.

\subsection{$Holog$ model}

The  derivations of the $Holog$ model by using the holographic approach can be found in Refs.~\cite{MZ19,MZ21}. 
Here we only give the formulas of Ref.~\cite{MZ21} which are used to calculate the cross sections for comparing  
with the JLab data and  making predictions for future experimental tests.

\begin{figure}[t]
\centering
\includegraphics[width=0.9\columnwidth,angle=0]{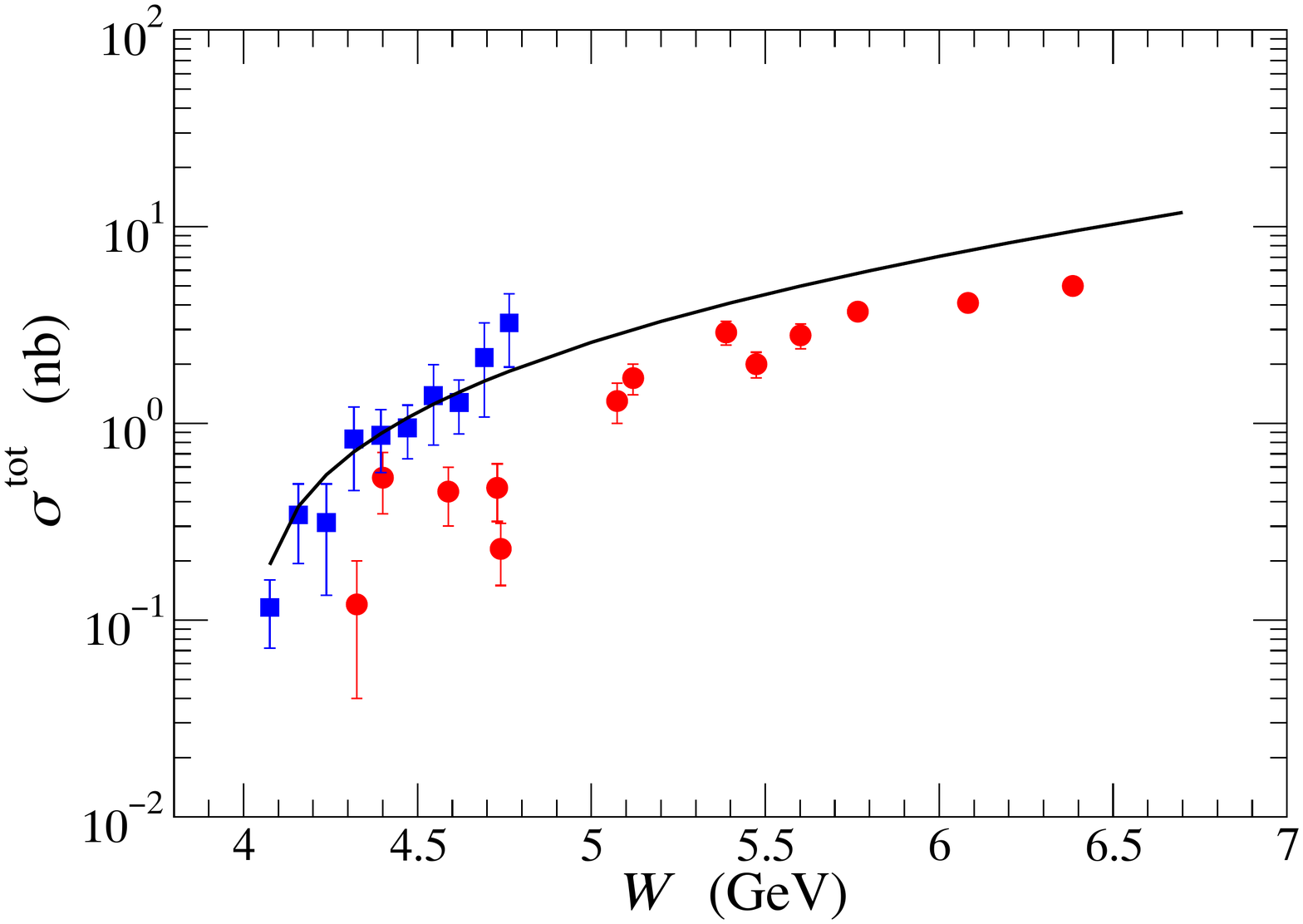}
\includegraphics[width=0.9\columnwidth,angle=0]{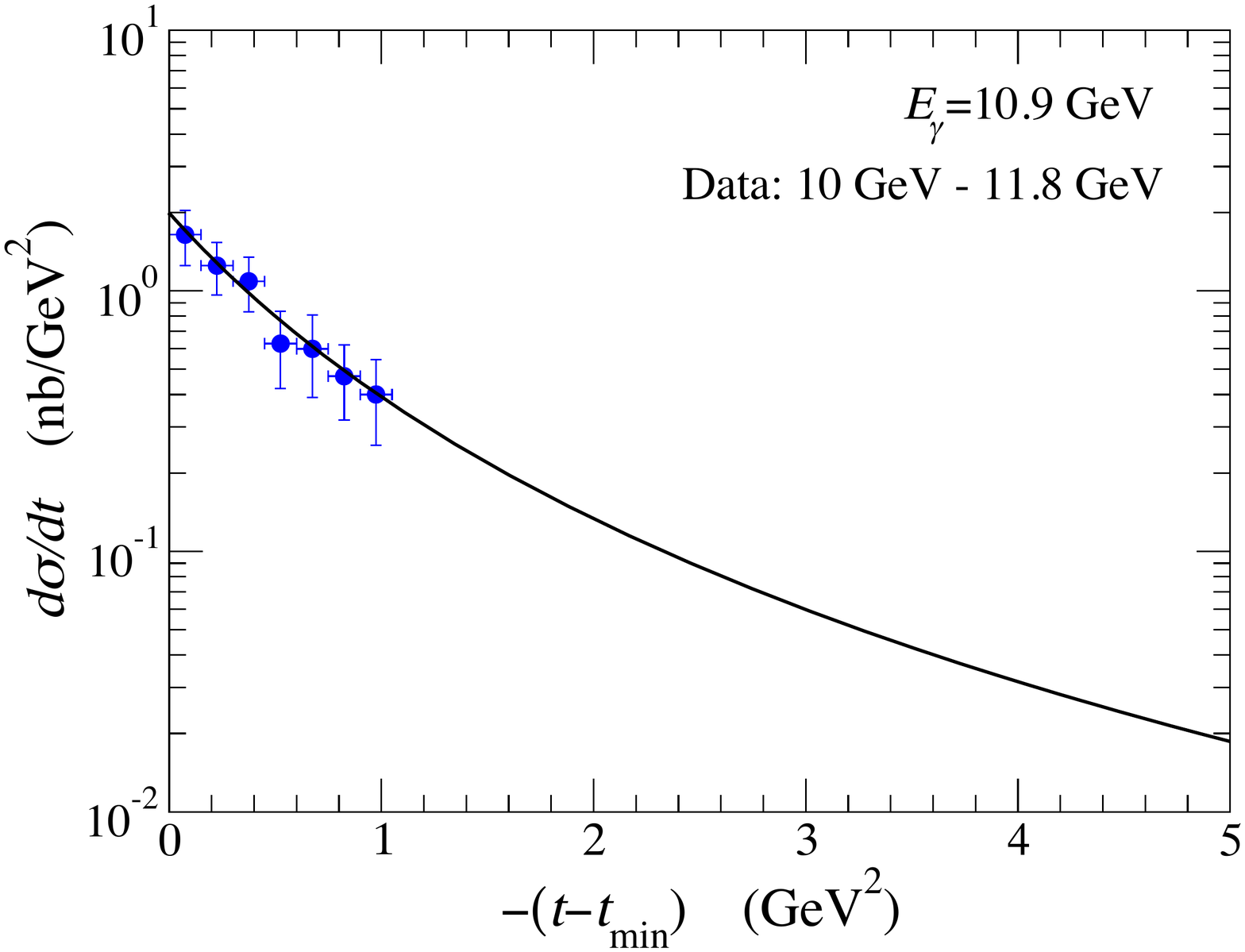}
\caption{$Holog$ model. Upper: total cross sections;
Lower: differential cross sections.}
\label{fig:tcrst-holo}
\end{figure}

Within the $Holog$ model, the $\gamma + N \rightarrow J/\Psi+N$ reaction is due to the exchange of scalar ($0^{++}$) 
and tensor ($2^{++}$) glueballs as illustrated in Fig.~\ref{fig:holo}. 
The resulting expression of the differential cross sections is of the following form: 
\begin{eqnarray}
\frac{d\sigma}{dt} &=& \left[A^2(0)\times\tilde{N} \right] I^2\, N_T
\left(\frac{s}{\tilde{\kappa}^2_N} \right)^2
\left(-\frac{t}{4m^2_N}+1 \right)
\nonumber \\ && \mbox{} \times
\left[\frac{1}{(1-{t}/{\Lambda_t^2})^2} \right]^2 ,
\end{eqnarray}
where $\Lambda_t=1.124 $ GeV, and
\begin{eqnarray}
I&=&\frac{3}{2}\frac{g_5f_{J/\Psi}}{M_{J/\Psi}} 
\nonumber \\ && \mbox{} \times
\left[\frac{1}{ \left(\frac{Q^2}{4\tilde{\kappa}^2_{J/\Psi}}+3\right)
\left(\frac{Q^2}{4\tilde{\kappa}^2_{J/\Psi}}+2\right)
\left(\frac{Q^2}{4\tilde{\kappa}^2_{J/\Psi}}+1\right)}\right ]_{Q^2=0} ,
 \\ 
N_T
&=&\frac{1}{64\pi} \left[e^2\frac{(2\kappa^2)^2}{g^2_5}\right]
\frac{\tilde{\kappa}^4_N}{\tilde{\kappa}^8_{J/\Psi}}
\end{eqnarray}
with
\begin{eqnarray}
g_5&=&\frac{2^{3/2}\tilde{\kappa}^2_{J/\Psi}}{f_{J/\Psi}m_{J/\Psi}} ,\\
\kappa&=&\sqrt{4\pi^2/N^2_c}.
\end{eqnarray}
The  parameters in the above equations are: $N_c=3$, $e=0.3$, $\tilde{\kappa}_N=0.35$, $m_N=0.94$~GeV, 
$m_{J/\Psi}=3.1$~GeV, $f_{J/\Psi}=0.405$~GeV, and
$\tilde{\kappa}_{J/\Psi}=1.03784$~GeV.
The normalization constant $[A^2(0)\times\tilde{N}]$ is fixed by fitting the JLab data, 
\begin{eqnarray}
A^2(0)\times\tilde{N} =43.11~\mu b/\mbox{GeV}^2.
\end{eqnarray}
The  total cross section is then  obtained by
\begin{eqnarray}
\sigma^{\rm tot}(s)=\int_{|t|_{\rm min}}^{|t|_{\rm max}} dt \frac{d\sigma}{dt} .
\end{eqnarray}
The results are compared with the JLab data in Fig.~\ref{fig:tcrst-holo}.

\section{ $N^*$ excitation }
\label{sec:N*}

To study the $N^*$($P_c$) in $\gamma + p \rightarrow J/\Psi+p$, we first note that there are coupled-channel effects 
due to the couplings between $J/\Psi N$,  $\rho N$, $\pi N$ and multi-meson $XN$ channels.
These coupled-channel effects were studied in Ref.~\cite{WL13} and were found to be small. 
Hence Eqs.~(\ref{eq:nstar-1}) and (\ref{eq:nstar-2}) are sufficient for the calculations presented below.

Since the width of $N^*$($P_c$) is very narrow $\sim 20 $~MeV, we can assume $M_{N^*} + \Sigma_{N^*}(E) \sim M_{R} 
- \frac{i}{2}\Gamma^{\rm tot}(W)$, and Eq.~(\ref{eq:nstar-2}) then becomes the usual Breit-Wigner form.
In the CM system, the amplitude of $\gamma(\mathbf{q},\lambda_\gamma) + N (-\mathbf{q},\lambda_N) \to N^*(J, M_J) \to
J/\Psi(\mathbf{k},m_{J/\Psi}) + N(-\mathbf{k}, m_s')$ can then be written as 
\begin{eqnarray}
 \braket{ {\bf k}, m_{J/\Psi} \, m'_s \vert t^{N^*}_{JM_J} \vert {\bf q},\lambda_\gamma\, \lambda_N }
 &=& 
 \braket{ {\bf k}, m_{J/\Psi} \, m'_s \vert \bar{F}_{N^*,J/\Psi N} \vert J M_J }
 \nonumber \\ && \hskip -3cm \mbox{} \times
\frac{1}{W  - M_R +\frac{i}{2}\Gamma^{\rm tot}(W)} 
\braket{ JM_J \vert {F}^\dagger_{N^*,\gamma N} \vert {\bf q},\lambda_\gamma \, \lambda_N },
\label{eq:nstar-amp}
\end{eqnarray}
where $\lambda_\gamma$ and $\lambda_N$ are the helicities of the photon and the incoming nucleon, respectively, 
$m_{J/\Psi}$ and $m_s'$ are the spin projections of the ${J/\Psi}$ and the recoiled nucleon.
The spin and its projection of the intermediate $N^*$ are denoted by $J$ and $M_J$, respectively.

Following Ref.~\cite{MSL06}, the matrix element of the $\gamma N \to N^*$ transition is
\begin{eqnarray}
 \braket{ JM_J \vert {F}^\dagger_{N^*,\gamma N} \vert \mathbf{q},\lambda_\gamma \, \lambda_N} &=&
\delta_{\lambda, (\lambda_\gamma -\lambda_N)}
 \frac{1}{(2\pi)^{3/2}} \sqrt{\frac{m_N^{} q_{N^*}^{}}{E_N(q)}}
 \nonumber \\ && \mbox{} \times
 A_\lambda D^J_{\lambda, M_J} (\phi_q,\theta_q,-\phi_q) ,
\label{eq:g-gn}
\end{eqnarray}
where $A_\lambda$ is the helicity amplitude of the $\gamma N \rightarrow N^*$ excitation,
$q_{N^*}^{}$ and $q$ are determined by $M_R = q_{N^*} + E_N(q_{N^*})$ and $W =q + E_N(q)$,
respectively, and
\begin{equation}
D^J_{\lambda, M_J}(\phi_q,\theta_q,-\phi_q) = e^{i(\lambda-M_J)\phi}d^{J}_{\lambda,M_J}(\theta_q).
\end{equation}
Here $d^{J}_{\lambda,M_J}(\theta_q)$ is the Wigner $d$-function.

The matrix element of the $N^* \to J/\Psi N$ transition is parameterized~\cite{MSL06} as
\begin{eqnarray}
&& \braket{ {\bf k}, m_{J/\Psi} \, m'_s \vert \bar{F}_{N^*,J/\Psi N} \vert JM_J }
\nonumber \\
&=& \sum_{LS} \sum_{M_LM_S}
\braket{JM_J \vert LS M_LM_S} 
\braket{SM_S \vert 1 \textstyle\frac{1}{2}m_{J/\Psi} m_{s}} 
Y_{LM_L}(\hat{\mathbf{k}}) \nonumber \\
&& \times  
\frac{1}{(2\pi)^{3/2}}\frac{1}{\sqrt{2E_{J/\Psi}(k)}}\sqrt{\frac{m_N^{}}{E_N(k)}}
\sqrt{\frac{8\pi^2M_{R}}{m_N k}}G^J_{LS} \left( \frac{k}{k_{N^*}} \right)^L ,\nonumber \\
&&
\label{eq:g-vn}
\end{eqnarray}
where $k$ and $k_{N^*}$ are determined by $W = E_{J/\Psi}(k) + E_N(k)$ and $M_{R} = E_{J/\Psi}(k_{N^*} ) 
+ E_N(k_{N^*})$, and $L$ and $S$ are the orbital angular momentum and total spin of the $J/\Psi$-N system.
$\braket{ j\,m|j_1\,j_2\,m_1\,m_2}$ is the Clebsch-Gordan coefficient, and $Y_{LM_L}(\hat{\bf k})$ is the 
spherical harmonic function.
We follow the dynamical approach of Ref.~\cite{MSL06} to define the $W$-dependence of 
$\Gamma^{\rm tot}(W)$ in Eq.~(\ref{eq:nstar-amp}) as
\begin{eqnarray}
\Gamma^{\rm tot}(W)=\Gamma^{\rm tot}_0 \frac{\rho(k)}{\rho(k_{N^*})} \left( \frac{k}{k_{N^*}} \right)^{2L}
\left(\frac{\Lambda^2}{(k-k_{N^*})^2+\Lambda^2} \right)^{2L+4} ,
\nonumber \\
\end{eqnarray}
where  $\rho(k)=\pi kE_N(k)E_{J/\Psi}(k)/(E_N(k)+E_{J/\Psi}(k))$ and $\Lambda=650$~MeV is a cutoff parameter.

The resonance amplitude of Eq.~(\ref{eq:nstar-amp}) then have parameters: the total decay width $\Gamma^{\rm tot}_0$, 
helicity amplitude $A_{\lambda}$ for $N^*\rightarrow \gamma N$ of Eq.~(\ref{eq:g-gn}), and $G_{LS}$ 
for $N^*\rightarrow  J/\Psi N$ of Eq.~(\ref{eq:g-vn}). 
These parameters can be related  to the partial decay widths defined by
\begin{eqnarray}
\Gamma_{N^*,\gamma N} &=& \int d{\bf q}  \delta ( M_{R} -E_N({\bf q})-|{\bf q}| ) 
\nonumber \\ && \mbox{} \times
\frac{2\pi}{2J+1}  \sum_{M_J}\sum_{\lambda_\gamma\, \lambda_N}
| \braket{ {\bf q},\lambda_\gamma\, \lambda_N | \bar{F}^\dagger_{N^*,\gamma N} | J M_J}|^2\
\nonumber \\
&=&\frac{q^2_{N^*}}{4\pi} \frac{m_N}{M_{R}} \frac{8}{2J+1}
( | A_{1/2}|^2 + |A_{3/2}|^2 ) ,
\label{eq:pwd-1} 
\end{eqnarray}
and
\begin{eqnarray}
\Gamma_{N^*,J/\Psi N} &=& \int d{\bf k} \delta ( M_{R} -E_{J/\Psi}({\bf k})-E_N({\bf k})) 
\nonumber \\ && \hskip -0.5cm  \mbox{} \times
\frac{2\pi}{2J+1} \sum_{M_J}\sum_{m_{J/\Psi} \, m'_{s}}
| \braket{{\bf k}, m_{J/\Psi}\, m'_s |\bar{F}_{N^*,J/\Psi N} | JM_J }|^2 
\nonumber \\
&=&\sum_{LS}| G^J_{LS}|^2 .
\label{eq:pwd-2}
\end{eqnarray}

Our objective here is to investigate the extent to which the available JLab data can be related to the $P_c(4337)$ state 
reported recently by the LHCb Collaboration~\cite{LHCb-21a}. 
We will assume that the non-resonant amplitude in Eq.~(\ref{eq:nstar-1}) is defined by the amplitude generated from 
either the $Pom-$pot or $2g+3g$ models described in the previous sections. 
The spin-parity of $P_c(4337)$ state was not well determined and has four possible specifications. 
For our illustrative purposes, it is sufficient to only consider $J^\pi(LS)=\frac{1}{2}^+(0,1), \frac{3}{2}^+(2,1)$, and
use their values of decay widths $\Gamma^{\rm tot}=29$ MeV and $\Gamma_{N^*,J/\psi N}=6.8$~MeV.
For simplicity, we set $A_{3/2}=0$ and hence $A_{1/2}$ is the only parameter in fitting the JLab data.

\begin{figure}[t]
\centering
\includegraphics[width=0.9\columnwidth,angle=0]{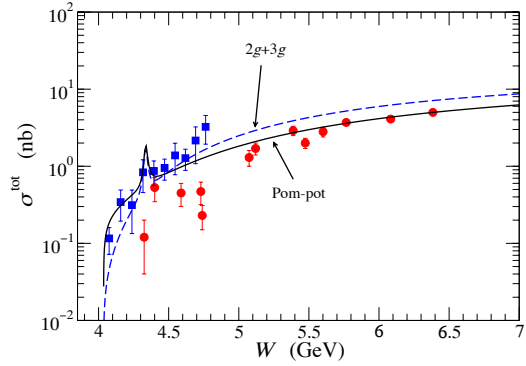}
\caption{Fits to the total cross section data of $\gamma+p\rightarrow J/\Psi+p$. 
The $P_c(4337)$ ($J^\pi(LS)=\frac{1}{2}^+(0,1)$ , $A_{1/2}=1\times 10^{-3}$ GeV$^{-2}$) is included in the fits 
with the non-resonant amplitudes calculated from either the $Pom$-pot (solid curve) or $2g+3g$ (dashed curve) 
models.
}
\label{fig:p-totcrst-pm-2g3g-nstar-1p}
\end{figure}

We first consider the specification $J^\pi(LS)=\frac{1}{2}^+(0,1)$. We find that the  total cross section calculated  
with $A_{1/2} \lesssim 1\times 10^{-3}$~GeV$^{-2}$ are within the uncertainties of the available data.
Our results with $A_{1/2}= 1\times 10^{-3}$~GeV$^{-2}$ and $A_{3/2}=0$ are shown in Fig.~\ref{fig:p-totcrst-pm-2g3g-nstar-1p}.
At the resonance  energy $W=M_R=4.337$~GeV, the results from $Pom$-pot and $2g-3g$ models are almost the same, 
indicating that our determination of $A_{1/2}$ is  rather model independent.
It appears that the available JLab data can not unambiguously exclude or confirm the $P_c(4337)$ state of LHCb.

\begin{figure}[t]
\centering
\includegraphics[width=0.9\columnwidth,angle=0]{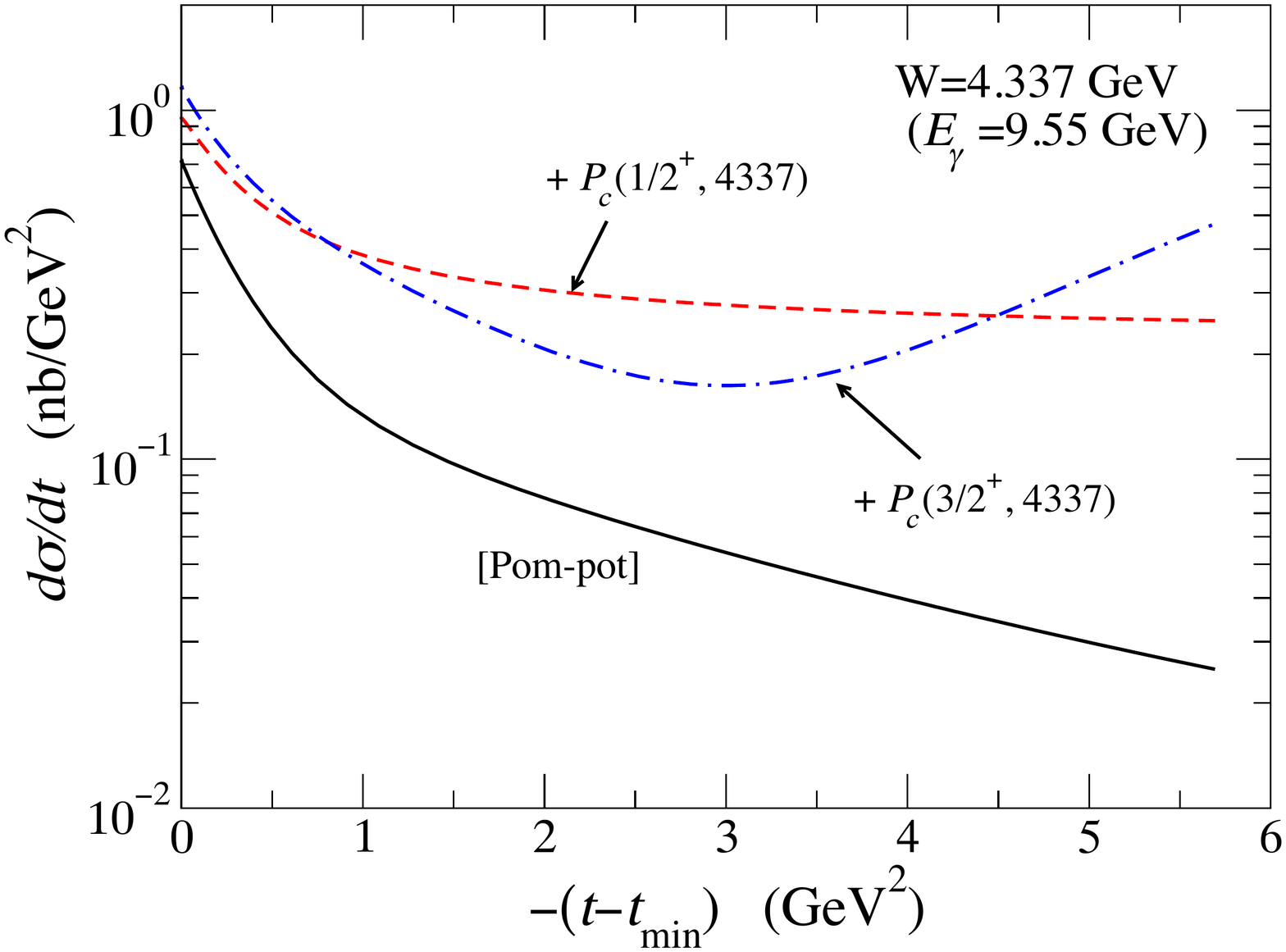}
\includegraphics[width=0.9\columnwidth,angle=0]{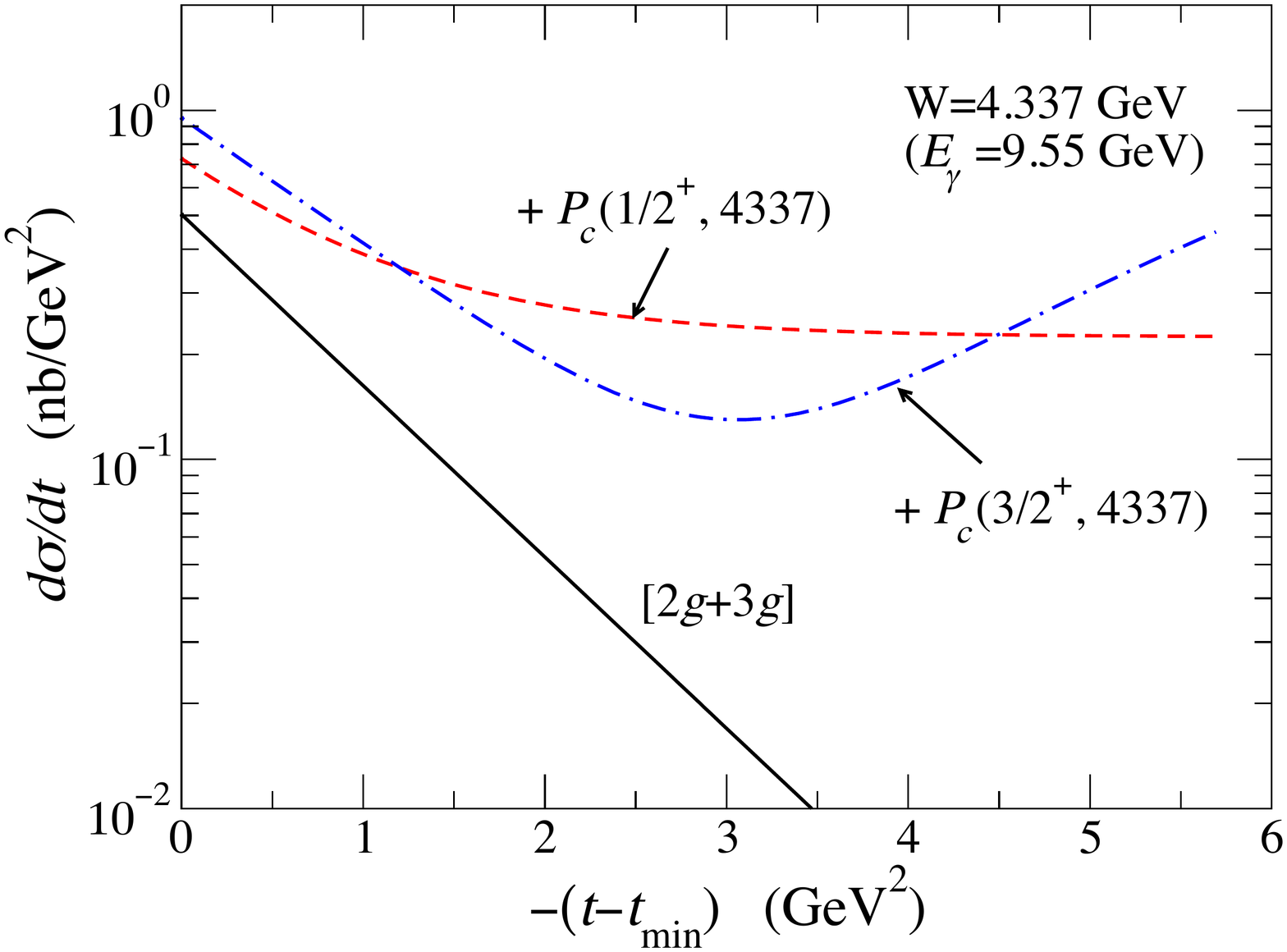}
\caption{The effects  of the resonance $P_b(4337)$ on the differential cross sections of $\gamma+p\rightarrow J/\Psi+p$.
The non-resonant amplitudes are calculated from $Pom$-pot (Upper) or $2g+3g$ (Lower) models. 
The  results from adding the  resonance amplitude are the curves indicated by $+P_c(\frac{1}{2}^+, 4337)$ and  
$+P_c(\frac{3}{2}^+,4337)$.
}
\label{fig:dsdt-pm-1p-3p}
\end{figure}

We next consider $J^\pi(LS)=\frac{3}{2}^+(2,1)$. 
We find that by also setting the helicity amplitude $A_{1/2}=1\times 10^{-3}$~GeV$^{-2}$ and $A_{3/2}=0$, 
the total cross sections are almost indistinguishable with that of $J^\pi(LS)=\frac{1}{2}^+(0,1)$.
However their differential cross sections are rather different at large $|t|$, as shown in Fig.~\ref{fig:dsdt-pm-1p-3p}. 
The cross sections for $J^\pi(LS)=\frac{3}{2}^+(2,1)$ raise at large $t$ can be a clear signal for identifying this $N^*$ state.

\section{$Pom$-CQM model}
\label{sec:pom-cqm}

In this section we describe the $Pom$-CQM model which can be used to relate the $J/\Psi$-N  potentials extracted from 
LQCD calculation~\cite{KS10b,sasaki-1} to the $\gamma + p \rightarrow J/\Psi  + p$ cross sections.
It is obtained by removing the VMD assumption within the $Pom$-pot model and using the CQM
to evaluate the $c\bar{c}$-loop mechanisms, illustrated in the upper part of Fig.~\ref{fig:fig-pom-dl-loop}.
Since the resulting model is also applicable to heavy quark system $\Upsilon$, we will use the notation $V$ for $J/\Psi$ in this section.

We start with the following Hamiltonian,
\begin{eqnarray}
H=H_0 +H_{\rm em}+v_{c\bar{c}} +v_{cN}\,,
\label{eq:h}
\end{eqnarray}
where $H_0$ is the free Hamiltonian, $v_{c\bar{c}}$ is a quark-quark potential of CQM, $v_{cN}$ is a quark-nucleon potential, and 
the electromagnetic $\gamma \rightarrow e^+e^-$, $c\bar{c}$ couplings are defined by
\begin{eqnarray}
H_{\rm em} &=& e\int dx \bar{\psi}_e(x)\gamma^\mu\psi_e(x) A_\mu(x)
\nonumber \\ && \mbox{}
+ e_c \int  dx\bar{\psi}_c(x)\gamma^\mu\psi_c(x)A_\mu(x) ,
\label{eq:h-em}
\end{eqnarray}
where $\psi_e(x)$ and $\psi_c(x)$ are the field operators of the electron with charge $e$ and the charm quark with charge 
$e_c$, respectively.

With the quark-quark potential $v_{c\bar{c}}$ in the Hamiltonian of Eq.~(\ref{eq:h}), we can generate the $J/\Psi$ wavefunction  
$\ket{\phi_{V}}$ in the rest frame of $J/\Psi$ by solving the following bound-state equation,
\begin{eqnarray}
(H_0+v_{c\bar{c}}) \ket{\phi_{V}} = E_V \ket{\phi_{V}} ,
\label{eq:wf-v}
\end{eqnarray}
where $E_V$ is the mass of $J/\Psi$.
The wavefunction $\phi_{V}$ can then be used to define $J/\Psi$-N potential $V_{VN,VN}$ from $v_{cN}$ by using the folding  
procedure~\cite{Feshbach92}:
\begin{eqnarray}
V_{VN,VN}= \braket{\phi_V, N|\sum_{c}v_{cN}|\phi_V,N} .
\label{eq:v-vnvn}
\end{eqnarray}
The $J/\Psi$ photo-production is also defined by the quark-nucleon potential $v_{cN}$ as
\begin{eqnarray}
B_{VN,\gamma N} &=&  
\bra{ \phi_V,N } \left[\sum_{c} v_{cN} \right] \,\frac{\ket{c\bar{c}} \bra{c\bar{c}}}{E_{c\bar{c}}-H_0}\, 
f_{\gamma, c\bar{c}} \ket{\gamma,N}  ,
\label{eq:photo-b}
\end{eqnarray}
where $f_{\gamma,c\bar{c}}$ is the $\gamma\rightarrow c\bar{c}$ coupling defined by the Hamiltonian $H_{\rm em}$ 
of Eq.~(\ref{eq:h-em}), and $E_{c\bar{c}}$ is the energy available to the propagation of $c\bar{c}$.

The unitarity condition requires the final state interaction (FSI) of the outgoing $VN$ and hence the total 
$\gamma+N\rightarrow J/\Psi +N$ amplitude defined by the Hamiltonian of Eqs.~(\ref{eq:h}) and (\ref{eq:h-em})  is 
\begin{eqnarray}
t^{\rm CQM}_{VN,\gamma N}(W) = B_{VN,\gamma N} + t^{\rm (fsi)}_{VN,\gamma N}(W) ,
\label{eq:t-cqm-0}
\end{eqnarray}
where
\begin{eqnarray}
t^{\rm (fsi)}_{VN,\gamma N}(W)=T_{VN,VN}(W)\frac{1}{W-H_0+i\epsilon}B_{VN,\gamma N} .
\label{eq:fsi-term}
\end{eqnarray}
Here $T_{VN,VN}(W)$ is the $VN\rightarrow VN$ scattering amplitude calculated from $V_{VN,VN}$ defined in 
Eq.~(\ref{eq:v-vnvn}) by solving the following Lippmann-Schwinger equation,
\begin{eqnarray}
T_{VN,VN}(W)= V_{VN,VN}+V_{VN,VN} \frac{1}{W-H_0+i\epsilon} T_{VN,VN}(W) .
\label{eq:lseq-0}
\end{eqnarray}

In order to fit the data from threshold to $W=300$~GeV, the Pomeron-exchange amplitude is included to define the total 
amplitude of $Pom$-CQM model as
\begin{eqnarray}
T_{VN,\gamma N}(W)= t^{\rm Pom}_{VN,\gamma N}(W)+t^{\rm CQM}_{VN,\gamma N}(W),
\label{eq:pom-cqm-t00}
\end{eqnarray} 
where $t^{\rm Pom}_{VN,\gamma N}(W)$ has been given  in Sec.~\ref{sec:Pom}.
To be consistent, we should calculate $t^{\rm Pom}_{VN,\gamma N}$ by also using CQM to calculate the $c\bar{c}$-loop  
illustrated in the lower part of Fig.~\ref{fig:fig-pom-dl-loop}.
This will require adjusting the Pomeron-quark parameters to fit the high energy  data, as shown in Fig.~\ref{fig:totcrst-all-v}, 
and is beyond the scope of this work.
Taking into account the spin indices and momentum  variables,  we now  give detailed formula for calculating 
the above equations in the following subsections.

\subsection{Determination of the $J/\Psi$ wavefunction}

For this exploratory study, we will not generate the $J/\Psi$ wavefunction from solving Eq.~(\ref{eq:wf-v}) within a CQM.
Instead we will assume a simple $s$-wave wavefunction,
\begin{eqnarray}
\phi_{{\bf p_V}, M_V}({\bf k}_1\, m_{s_q}, {\bf k}_2\, m_{s_{\bar{q}}})
&=& \delta({\bf p}_V-{\bf k}_1-{\bf k}_2) 
\nonumber \\ && \mbox{} \times \textstyle
\braket{J_VM_V|\frac{1}{2}\frac{1}{2}m_{s_q}m_{s_{\bar{q}}}}
\phi({\bf k})\,,
\label{eq:phik}
\end{eqnarray}
where ${\bf p}_V$ is the momentum of $J/\Psi$, ${\bf k}_i$ is the momentum of the $i$-th  quark,
${\bf k}=\frac{{\bf k}_1-{\bf k}_2}{2}$, $J_VM_V$ is the total angular momentum of $J/\Psi$, $m_{s_q}$ and $m_{s_{\bar{q}}}$
are  the spins of quarks $c$ and $\bar{c}$, respectively.
We assume a simple Gaussian form  
\begin{eqnarray}
\phi({\bf k})=N_0\,e^{-{\bf k}^2/b^2},
\label{eq:phi-exp}
\end{eqnarray}
where $N_0$ is determined by the normalization condition, 
\begin{eqnarray}
\int d{\bf k}\, \phi^*({\bf k})\phi({\bf k})=1 .
\end{eqnarray}

\begin{figure}[t]
\centering
\includegraphics[width=0.9\columnwidth,angle=0]{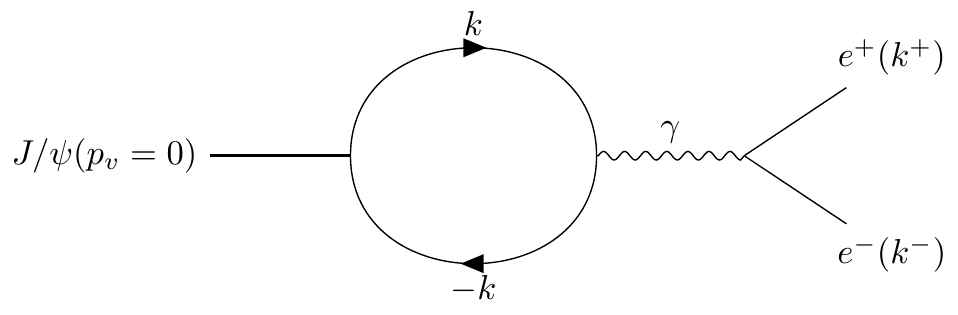}
\caption{ Momentum variables of $J/\Psi\rightarrow  e^+e^-$ decay in the rest frame of $J/\Psi$.
}
\label{fig:jpsi-decay}
\end{figure}

The amplitude of $J/\Psi\rightarrow c({\bf k})+\bar{c}(-{\bf k})\rightarrow \gamma \rightarrow  e^+({\bf k}_e)+e^-(-{\bf k}_e)$ 
in the rest frame of $J/\Psi$ with  ${\bf p_V}=0$, as illustrated in Fig.~\ref{fig:jpsi-decay}, is of the following form:
\begin{eqnarray}
\Braket{k_{+}m_{s_+}k_{-}m_{s_-}|T|\phi_{{\bf p_V},M_V}} &=& \frac{e}{(2\pi)^3}
\left[ \bar{u}_{m_{s_{+}}}({\bf k}_e)\gamma_\mu\,v_{m_{s_{-}}}(-{\bf k}_e) \right]
\nonumber \\ && \mbox{} \times
\frac{1}{m^2_{V}}F^\mu_{M_V} ,
\end{eqnarray}
where $k_+=(E_e({\bf k}_e),{\bf k}_e)$, $k_-=(E_e({\bf k}_e),-{\bf k}_e)$, and
\begin{eqnarray}
F^\mu_{M_V}&=e_c& \sum_{m_{s_q},m_{s_{\bar{q}}}} \int d{\bf k}
\frac{m_c}{E_c({\bf k})}
\left[ \bar{v}_{m_{s_{\bar{q}}}}({\bf k})\gamma^\mu\, u_{m_{s_{q}}}(-{\bf k}) \right]
\nonumber \\ && \mbox{} \times \textstyle
\braket{J_VM_V|\frac{1}{2}\frac{1}{2}m_{s_q}m_{s_{\bar{q}}} }
\phi({\bf k}) .
\label{eq:fmu}
\end{eqnarray}
Here $e_c=\frac{2}{3}e$ is the charge of $c$ quark.
The decay width of $J/\Psi\rightarrow e^+e^-$ is then calculated from
\begin{eqnarray}
\Gamma_{V\rightarrow  e^+e^-}&=&(2\pi)\int d{\bf k}_e\,\delta(m_V-2E_e({\bf k}_e))
\nonumber \\ && \mbox{} \times
\sum_{m_{s_+}m_{s_-}}
\frac{1}{2J_V+1}\sum_{M_V} | \braket{k_{+}m_{s_+}k_{-}m_{s_-}|T|\phi_{{\bf p_V},M_V} } |^2 
\nonumber \\ &=&
\frac{k_e}{(2\pi)^5}\frac{1}{m^5_V} \left( \frac{2}{3}e^2 \right)^2
\nonumber \\ && \mbox{} \times 
\frac{1}{2J_V+1} \sum_{M_V}A^{\mu\nu}(k_+,k_-)  F_{M_V,\mu} F^*_{M_v,\nu} 
\label{eq:ewidth-1}
\end{eqnarray}
with
\begin{eqnarray}
A^{\mu\nu}(k_+,k_-) &=&
k^\mu_+k^\nu_-+k^\nu_+k^\mu_- -g^{\mu\nu}(k_+\cdot k_-+m^2_e)\,.
\end{eqnarray}
With the above formulas, the parameter $b$ of the $J/\Psi$ wavefunction defined by Eqs.~(\ref{eq:phik}) and (\ref{eq:phi-exp})
is determined by fitting the data of $J/\Psi \rightarrow e^+e^-$ decay. 
We find that the decay width $\Gamma_{J/\Psi \rightarrow e^+e^-}=5.1$~keV  can be reproduced with $b=2.5 $~GeV.

\subsection{$J/\Psi$-N potential}

To evaluate Eq.~(\ref{eq:v-vnvn}), we assume for simplicity that quark-nucleon interaction, $v_{cN}$, is independent
of spin variables, 
\begin{eqnarray}
\bra{{\bf k}\,m_{s_c}\,, {\bf p}\,m_{s_N}} v_{cN} \ket{{\bf  k}'\,m'_{s_c}\,, {\bf p}'\,m'_{s_N}}
&=&
\delta_{m_{s_c},m'_{s_c}}\delta_{m_{s_N},m'_{s_N}}
\nonumber \\ && \hskip -2cm \mbox{} \times
\delta({\bf k+p-k'-p'}) \braket{{\bf q}|v_{cN}|{\bf q'}} , 
\label{eq:vc-pot}
\end{eqnarray}
where the relative  momenta are  defined by
\begin{eqnarray}
{\bf q}&=& \frac{m_N{\bf k}-m_c{\bf p}}{m_N+m_c}, \label{eq:kin1}  \\
{\bf q}'&=& \frac{m_N{\bf k}'-m_c{\bf p}'}{m_N+m_c} . \label{eq:kin2} 
\end{eqnarray}
Here $m_c$ and $m_N$ are the masses of the quark $c$ and the nucleon, respectively.

\begin{figure}[t]
\centering
\includegraphics[width=0.9\columnwidth,angle=0]{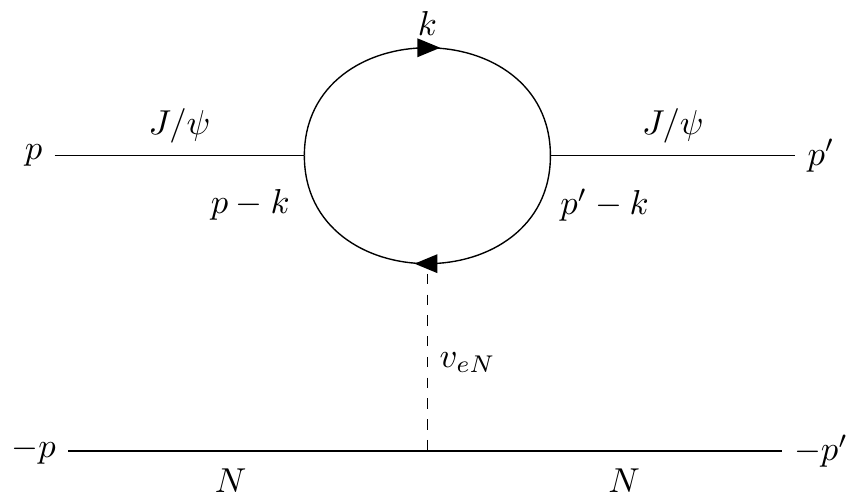}
\caption{  
Momentum variables for the calculation of $V_{VN,VN}$ defined by Eq.~(\ref{eq:vc-pot}). }
\label{fig:cqm-fig1}
\end{figure}

With the spin independent potential of Eq.~(\ref{eq:vc-pot}) and the $s$-wave $J/\Psi$ wavefucntion of Eq.~(\ref{eq:phik}),
the  $J/\Psi$-N defined by Eq.~(\ref{eq:v-vnvn}) is also spin independent.
In the CM frame, ${\bf k}_V= - {\bf p}$ and ${\bf k}'_V= - {\bf p}'$ as illustrated in Fig.~\ref{fig:cqm-fig1}, 
the matrix element of Eq.~(\ref{eq:v-vnvn}) can be written  as
\begin{eqnarray}
\braket{ {\bf k}_V m_{V}, {\bf  p} m_{s}|V_{VN,VN}|{\bf k}'_V m'_{V}, {\bf p}' m'_{s}}
&=& \delta_{m_{V},m'_{V}}\delta_{m_{s},m'_{s}} 
\nonumber \\  && \hskip -3cm \mbox{} \times
\delta({\bf k}_V+{\bf p}-{\bf k}'_V-{\bf p}')
\braket{{\bf p}|V_{VN}|{\bf p}' } ,
 \label{eq:vjpsin} 
\end{eqnarray}
where
\begin{eqnarray}
\braket{ {\bf p}|V_{VN}|{\bf p}'} &=& 2 \int  d{\bf k} \,  \textstyle \phi^* \left({\bf k}-\frac{{\bf p}}{2} \right)
\nonumber \\ && \mbox{} \times
\Braket{{\bf p}-\frac{ m_N}{m_N+m_c}{\bf k}|v_{cN}|{\bf p}'-\frac{ m_N}{m_N+m_c}{\bf k}}
\nonumber \\ && \mbox{} \times
\textstyle \phi({\bf k}-\frac{{\bf p}'}{2}) . 
\label{eq:v-vnvn-0}
\end{eqnarray}
Here the factor $2$ is from summing  the contributions from two quarks in $J/\Psi$, and we have used the definitions of
Eqs.~(\ref{eq:kin1}) and (\ref{eq:kin2}).

For simplicity, we choose a Yukawa form $v_{cN}(r)=\alpha \frac{e^{-\mu r}}{r}$ with $r$ being the distance between two quarks.
We then have 
\begin{eqnarray}
\Braket{ {\bf p}-\frac{ m_N}{m_N+m_c}{\bf k}|v_{cN}|{\bf p}'-\frac{ m_N}{m_N+m_c}{\bf k} }
= v_{cN}({\bf p} - {\bf p}') ,
\label{eq:yukawa}
\end{eqnarray}
where
\begin{eqnarray}
v_{cN}({\bf p-p'})&=&\frac{\alpha}{(2\pi)^3}\int\, d{\bf r}\,
e^{i{\bf (p-p')}\cdot {\bf r}} \left( \frac{\alpha e^{-\mu r}}{r} \right)
\nonumber \\
&=&  \frac{2\alpha}{(2\pi)^2}\frac{1}{ {\bf (p-p')}^2+\mu^2 } .
\end{eqnarray}
By using Eq.~(\ref{eq:yukawa}), Eq.~(\ref{eq:v-vnvn-0}) becomes the following factorized form,
\begin{eqnarray}
\braket{ {\bf p}|V_{VN}|{\bf p}' } &=& F_V({\bf t}) \left[ 2v_{cN}({\bf t}) \right] ,
\label{eq:v-vn-f}
\end{eqnarray}
where ${\bf t}={\bf p}-{\bf  p}'$ and
\begin{eqnarray}
F_V({\bf t}) &=& \int  d{\bf k} \, \textstyle
\phi^*({\bf k}- \frac{{\bf p}}{2}) \phi({\bf k}-\frac{{\bf p}'}{2}) \nonumber \\
&=&\int  d{\bf k} \,\textstyle
\phi^*({\bf k}-\frac{{\bf t}}{2}) \phi({\bf k})
\label{eq:v-ff}
\end{eqnarray}
is the form factor of the vector meson $V$.

The $J/\Psi$-N potential defined in Eqs.~(\ref{eq:vjpsin})-(\ref{eq:v-ff}) is then used to get scattering amplitude
by solving Eq.~(\ref{eq:lseq-0}). 
The resulting amplitude is spin independent and is 
\begin{eqnarray}
\Braket{ {\bf p}\, m_V m_{s}|T_{VN,VN}(W)|{\bf p}' \, m'_V m'_{s} }
&=& \delta_{m_V,m'_V}\delta_{m_s,m'_s} 
\nonumber \\ && \hskip -0.5cm \mbox{} \times
\Braket{ {\bf p}^{'}|T_{VN}(W)|{\bf p} },
\label{eq:t-vnvn-0}
\end{eqnarray}
where  $\braket{ {\bf p}'|T_{VN}((W)|{\bf p} }$ is from solving
\begin{eqnarray}
\braket{{\bf p}'|T_{VN}(W)|{\bf p}} &=& \braket{ {\bf p}'|V_{VN}|{\bf p} } 
\nonumber \\ && \mbox{} \hskip -2.cm
+ \int d{\bf p^{''}} \braket{{\bf p}'|V_{VN}|{\bf p}^{''}}
\frac{1}{W- E_N(p^{''})-E_V(p^{''})+i\epsilon}
\nonumber \\ && \mbox{} \hskip -1.5cm \times
\braket{{\bf p}^{''}|T_{VN}(W)|{\bf p}} .
\label{eq:lseq-00}
\end{eqnarray}

\subsection{$J/\Psi$ photo-production }

\begin{figure}[t]
\centering
\includegraphics[width=0.9\columnwidth,angle=0]{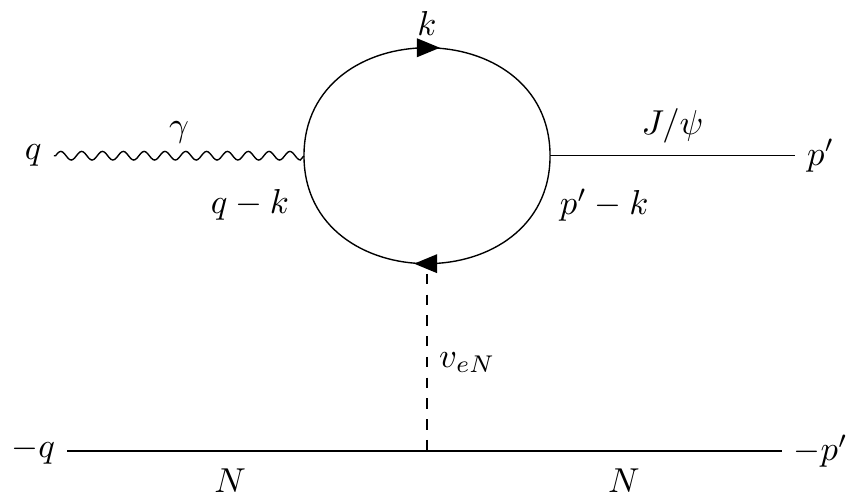}
\caption{Momentum variables for calculating the matrix element of photo-production of Eq.~(\ref{eq:t-gnvn-0}).  }
\label{fig:cqm-fig2}
\end{figure}

By using Eq.~(\ref{eq:phik}) for the $J/\Psi$ wave function  and  Eq.~(\ref{eq:vc-pot}) for quark-N potential, the matrix element 
of photo-production of Eq.~(\ref{eq:photo-b}) can be calculated.
With the variables in the CM system, as illustrated in Fig.~\ref{fig:cqm-fig2}, we follow the dynamical formulation of 
Refs.~\cite{SL96,MSL06,JLMS07,KNLS13} to write 
\begin{eqnarray}
&& \Braket{ {\bf p}'m_V m_{s'}|B_{VN,\gamma N}|{\bf q}\lambda m_{s}}
\nonumber \\
&=&
\sum_{m_c,m_{\bar{c}}}\frac{1}{(2\pi)^3}\frac{e_c}{\sqrt{2|{\bf q}|}}
\int d{\bf k} \textstyle
\braket{J_Vm_V|\frac{1}{2}\frac{1}{2}m_c,m_{\bar{c}} } \phi({\bf k}-\frac{1}{2}{\bf p}')
\nonumber \\ && \mbox{} \times
\delta_{m_{s},m_{s'}}  
\Braket{{\bf p}'-\frac{ m_N}{m_N+m_c}{\bf k}|v_{cN}|{\bf q}-\frac{ m_N}{m_N+m_c}{\bf k} }
\nonumber \\ && \mbox{} \times
\frac{1}{W-E_N({\bf q})-E_c({\bf q-k})-E_c({\bf k})+i\epsilon} 
\nonumber  \\ && \mbox{} \times 
\bar{u}_{m_q}({\bf k})[{\bf \epsilon_\lambda}\cdot {\bf \gamma}] v_{m_{\bar{q}}}({\bf q-k}) .
\label{eq:t-gnvn-0}
\end{eqnarray}
If we choose the same Yukawa form of $v_{cN}$ of Eq.~(\ref{eq:yukawa}), we obtain the following factorized form:
\begin{eqnarray}
\Braket{ {\bf p}' m_V m_{s'}|B_{VN,\gamma N}|{\bf q}\lambda m_{s} } &=&
C_{\lambda,m_V} \delta_{m_s,m_s'} B({\bf p}',{\bf q})
\nonumber \\ && \mbox{} \times
\left[ 2 v_{cN}({\bf q}-{\bf p}') \right] ,
\label{eq:mtx-b-1}
\end{eqnarray}
where
\begin{eqnarray}
 B({\bf p}',{\bf q})
&&=\frac{1}{(2\pi)^3}\frac{e_c}{\sqrt{2|{\bf q}|}}
\int d{\bf k} \, \phi({\bf k}-{\textstyle\frac{1}{2}}{\bf p}')
\nonumber \\ && \mbox{} \times
\frac{1}{W-E_N({\bf q})-E_c({\bf q-k})-E_c({\bf k})+i\epsilon} 
\nonumber  \\ && \mbox{} \times
\sqrt{\frac{E_c({\bf k})+m_c}{2E_c({\bf k})}}\sqrt{\frac{E_c({\bf q- k})+m_c}{2E_c({\bf q- k})}}
\nonumber  \\ && \mbox{} \times
\left(1-\frac{{\bf k}\cdot{\bf (q-k)}}{[E_c({\bf k})+m_c][E_c({\bf q- k})+m_c]} \right)
\label{eq:tmx-b-2}
\end{eqnarray}
and 
\begin{eqnarray}
C_{\lambda,m_V}=\sum_{m_q,m_{\bar{q}}} \textstyle
\braket{J_Vm_V|\frac{1}{2}\frac{1}{2}m_qm_{\bar{q}} }
\braket{m_{\bar{q}}|{\bf \sigma}\cdot {\bf \epsilon_\lambda}|m_q} .
\end{eqnarray}

With Eqs.~(\ref{eq:t-vnvn-0}) and (\ref{eq:t-gnvn-0}), the matrix element of Eq.~(\ref{eq:t-cqm-0}) can be calculated as
\begin{eqnarray}
\Braket{ {\bf p}' m_V m'_s | t^{\rm (CQM)}_{VN,\gamma N} | {\bf q} \lambda m_s}
&=&
\Braket{{\bf p}'m_Vm'_s|B_{VN,\gamma N}|{\bf q}\lambda m_s}
\nonumber \\ && \mbox{} \hskip -0.5cm
+ \braket{{\bf p}'m_Vm'_s|t^{\rm (fsi)}_{VN,\gamma N}|{\bf q}\lambda m_s} ,
\label{eq:pom-cqm-t0}
\end{eqnarray}
where
\begin{eqnarray}
&& \Braket{{\bf p}'m_Vm'_s|t^{\rm (fsi)}_{VN,\gamma N}|{\bf q}\lambda m_s}
\nonumber \\
&=&
\sum_{m^{''}_V,m^{''}_s} \int d{\bf  p}^{''} 
\Braket{{\bf p}\, m_Vm'_s|T_{VN,VN}|{\bf  p}^{''}m^{''}_V,m^{''}_s} 
\nonumber \\ && \mbox{} \times
\frac{1}{W-E_N(p^{''})-E_V(p^{''})+i\epsilon}
\nonumber \\ && \mbox{} \times
\Braket{{\bf p}^{''}m^{''}_Vm^{''}_s|B_{VN,\gamma N}|{\bf q}\lambda, m_s} . 
\label{eq:tfsi-mx}
\end{eqnarray}
The total amplitude of Eq.~(\ref{eq:pom-cqm-t00}) of the $Pom$-CQM model is then of the following form:
\begin{eqnarray}
\Braket{ {\bf p}'m_Vm'_s|T^{\rm Pom-CQM}_{VN,\gamma N}|{\bf q}\lambda, m_s}
&=& \Braket{{\bf p}'m_Vm'_s|t^{\rm Pom}_{VN,\gamma N}|{\bf q}\lambda, m_s}
\nonumber \\ && \hskip -1.5cm \mbox{} 
+ \Braket{{\bf p}m_Vm'_s|t^{\rm CQM}_{VN,\gamma N}|{\bf q}\lambda, m_s} ,
\label{eq:pom-cqm-amp}
\end{eqnarray}
where $t^{\rm Pom}_{VN,\gamma N}$ has been given in Sec.~\ref{sec:Pom}.

\subsection{Results}

We now turn to discussing the results from using Eq.~(\ref{eq:pom-cqm-amp}).
We first consider the case that $\Braket{{\bf p}'|V_{VN}|{\bf  p}}$ of Eq.~(\ref{eq:v-vn-f}) can be identified with
the matrix element of $J/\Psi$-N potential extracted from LQCD shown in Fig.~\ref{fig:v-sasaki-lqcd}.
It turns out that the form factor $F_V({\bf t})$ drops from $F_v(0)=1$ very slowly as momentum transfer $t$ 
increases, and hence we can use the approximation $\braket{{\bf p}|V_{VN}|{\bf p}'} \sim 2v_{cN}({\bf t})$ for 
Eq.~(\ref{eq:v-vn-f}).
Thus the parameters of $v_{cN}(r)$ can be directly determined by the potentials $pot$-1 and $pot$-2 
(Fig.~\ref{fig:v-sasaki-lqcd}) from LQCD.

\begin{figure}[t]
\centering
\includegraphics[width=0.9\columnwidth,angle=0]{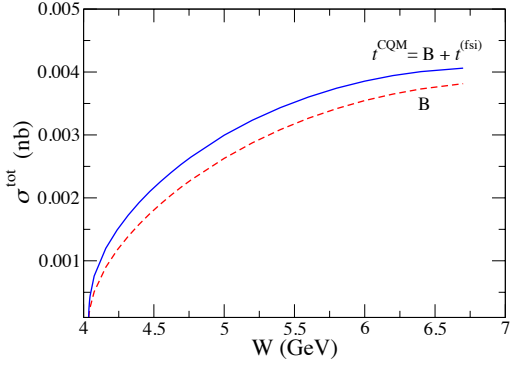}
\includegraphics[width=0.9\columnwidth,angle=0]{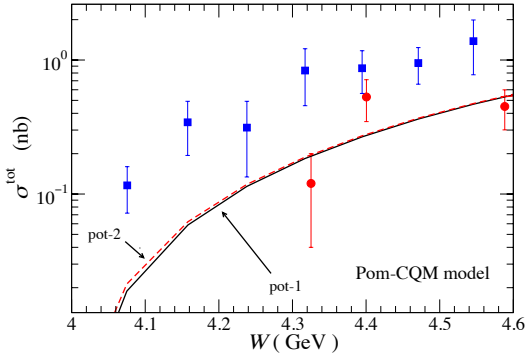}
\caption{$Pom$-CQM model. 
The results are from using the parameters of quark-nucleon potential $v_{cN}$ determined from the
$J/\Psi$-N potentials extracted from LQCD. 
Upper: effect of final-state interaction. 
Lower: Comparisons with JLab data. pot-1 (pot-2) indicates that the parameters of quark-nucleon potential 
$v_{cN}$ are determined by using the potentials pot-1 (pot-2) extracted from LQCD shown in Fig.~\ref{fig:v-sasaki-lqcd}. }
\label{fig:qmd-lqcd}
\end{figure}

We  find  that the FSI effects due to the $t^{\rm (fsi)}_{VN,VN}$ term in Eq.~(\ref{eq:pom-cqm-t0}) are significant 
in increasing the total cross sections, as shown in the upper part of Fig.~\ref{fig:qmd-lqcd}. 
However, the predicted magnitudes, from using the parameters of $v_{cN}$ determined by using $pot$-1 or $pot$-2 
in Fig.~\ref{fig:v-sasaki-lqcd}, are very small. 
When Pomeron-exchange is included, the JLab data can not be explained, as shown in the lower part of Fig.~\ref{fig:qmd-lqcd}.

\begin{figure}[t]
\centering
\includegraphics[width=0.9\columnwidth,angle=0]{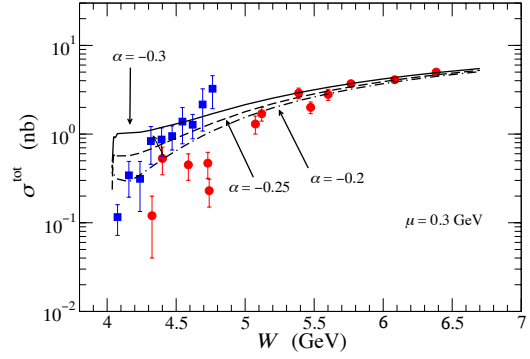}
\includegraphics[width=0.9\columnwidth,angle=0]{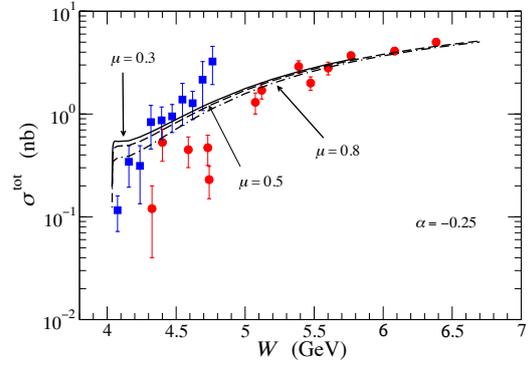}
\caption{Dependence of the total cross sections on the  parameter $\alpha$ (upper)
and $\beta$ (lower) of the quark-nucleon potential $v_{cN}=\alpha\frac{e^{-\mu r}}{r}$
within the  $Pom$-CQM model.  }
\label{fig:tcrst-pom-qmd}
\end{figure}

Since the LQCD calculation of Ref.~\cite{KS10b} was a very first step, the results shown in Fig.~\ref{fig:v-sasaki-lqcd} 
will certainly be revised by using more advanced LQCD calculations, such as those reported in Refs.~\cite{KS13, KS15a} 
for the spin-dependent $J/\Psi$-N potential and Ref.~\cite{LDHI22} for the $\phi$-N potential. 
Thus we will determine the parameters $\alpha$ and $\mu$ of quark-N potential $v_{cN}=\alpha\frac{e^{-\mu r}}{r}$  by  
fitting the JLab data. 
The resulting  parameters may be taken as the data to test future LQCD calculations.
In the upper part of Fig.~\ref{fig:tcrst-pom-qmd}, we show the sensitivity to the parameter $\alpha$ and $\alpha=-0.25$
clearly is favored by the data at low energies.
In the lower part of Fig.~\ref{fig:tcrst-pom-qmd}, we show the dependence on $\mu$ and $\mu=0.3$~GeV seems favored 
by the data.

\begin{figure}[t]
\centering
\includegraphics[width=0.9\columnwidth,angle=0]{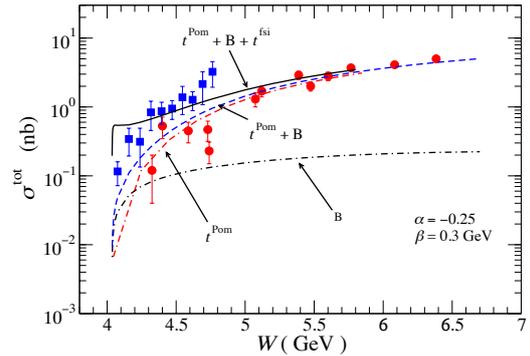}
\caption{Comparisons of the contribution from each amplitude in Eqs.~(\ref{eq:pom-cqm-t0})-(\ref{eq:pom-cqm-amp}). }
\label{fig:tcrst-pom-qmd-mech}
\end{figure}

\begin{figure}[t]
\centering
\includegraphics[width=0.9\columnwidth,angle=0]{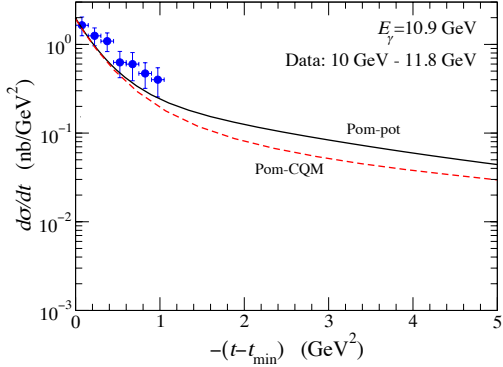}
\caption{Comparisons of differential cross sections from $Pom$-pot and $Pom$-CQM models. }
\label{fig:dsdt-pot-qmd}
\end{figure}

The best  fit to JLab data is from using $\alpha=-0.25$ and $\mu=0.3$ GeV.
The contributions from each mechanism are shown in  Fig.~\ref{fig:tcrst-pom-qmd-mech}.
The Born term $B_{VN,\gamma NN}$ (B) of $Pom$-CQM is larger than the Pomeron-exchange ($t^{\rm Pom}$)
near threshold.
When the Pomeron-exchange is included ($t^{\rm Pom}$+B), the predicted results are still below the Jlab data. 
Only when FSI effects are included ($t^{\rm Pom}+B+t^{\rm (fsi)}$), we obtain the black solid curve which agrees
reasonably with the Jlab data. 
The pronounced enhancement due to FSI at very near threshold is similar to that seen in nuclear reactions. 
The predicted differential cross sections are comparable to that of the  $Pom$-pot model in describing the Jab data, 
as shown in Fig.~\ref{fig:dsdt-pot-qmd}.

Apparently, the results presented here need to be examined by using the $J/\Psi$ wavefunction generated 
from a more realistic CQM which fits the spectrum of charmonium ($c\bar{c}$) and the width of
$J/\Psi\rightarrow e^+e^-$ decay. 
Since the momentum-transfer for the $\gamma\rightarrow c\bar{c}\rightarrow J/\Psi$ loop (Fig.~\ref{fig:cqm-fig1}) 
and $J/\Psi \rightarrow c\bar{c}\rightarrow J/\Psi$ loop (Fig.~\ref{fig:cqm-fig2}) are rather different
in the threshold region, the parameters for the quark-nucleon potential $v_{cN}$ may have  to be
determined differently for these two calculations. 
It is, of course, most important to have new calculations of $J/\Psi$-N interaction from LQCD and the DSE models 
which can be tested within this reaction model.

\section{Predictions}
\label{sec:predict}

\begin{figure}[t]
\centering
\includegraphics[width=0.9\columnwidth,angle=0]{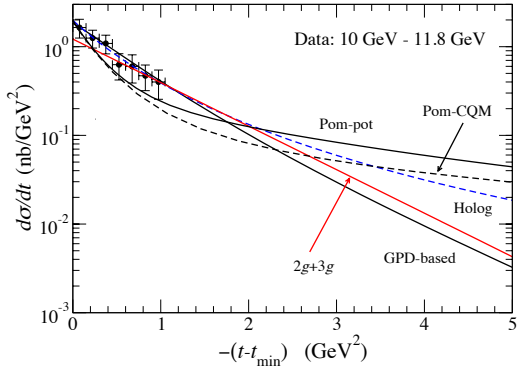}
\caption{Comparison of the differential cross sections from the five models presented in this paper.}
\label{fig:dsdt-all-model}
\end{figure}

\begin{figure}[t]
\centering
\includegraphics[width=0.9\columnwidth,angle=0]{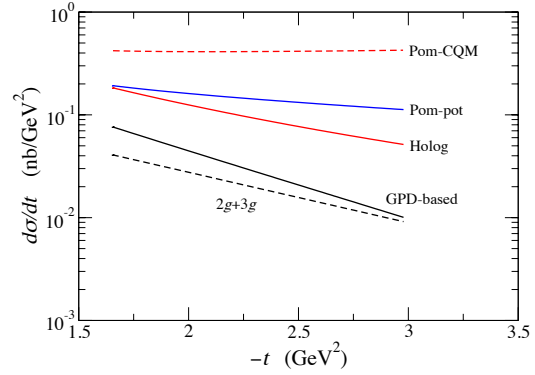}
\caption{Comparison of the differential cross sections at invariant mass $W=4.056$~GeV
predicted by  the five models presented in this paper.}
\label{fig:dsdt-all-model-4p056}
\end{figure}

The available JLab data can be reasonably fitted by the six models presented in this paper. 
Their differences can only be tested by using more extensive data from the future experiments~\cite{MJPC16,JLab-16,ATHENNA-12}.
This can be seen in Fig.~\ref{fig:dsdt-all-model} for the differential cross sections at large $|t|$.
Their differences become very dramatic at very near threshold, $W=4.056$~GeV, as shown in 
Fig.~\ref{fig:dsdt-all-model-4p056}.

\begin{figure}[t]
\centering
\includegraphics[width=0.9\columnwidth,angle=0]{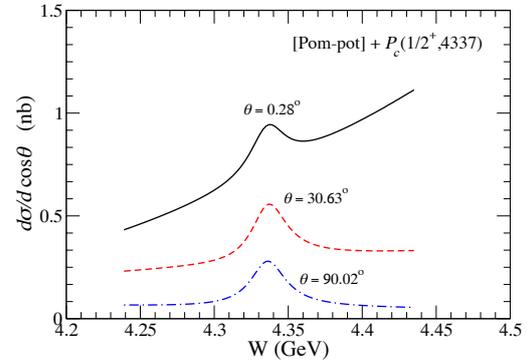}
\includegraphics[width=0.9\columnwidth,angle=0]{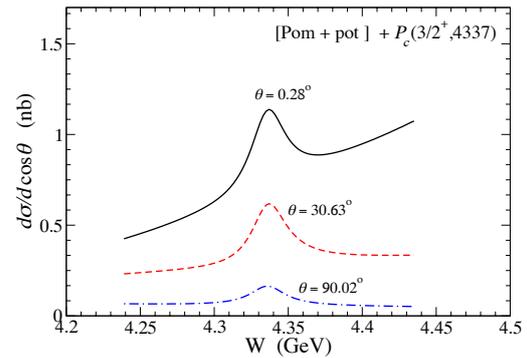}
\caption{The energy-dependence of the differential cross sections calculated from including
the nucleon resonance $P_b(4337)$ with $J^\pi = \frac12^+$ and $\frac32^+$. $\theta$ is the scattering angle of the outgoing
$J/\Psi$. 
}
\label{fig:dsdt-pm-w-1p-3p}
\end{figure}

For the search of $N^*$ states, an effective way is to examine the energy-dependence of the differential 
cross sections at a given scattering angle $\theta$.  
Our predictions for the $P_c(4337)$ state are given in Fig.~\ref{fig:dsdt-pm-w-1p-3p}. 
It will be interesting to see that these predictions and the $t$-dependence of differential cross sections, 
shown in Fig.~\ref{fig:dsdt-pm-1p-3p}, can be tested by the forthcoming data from JLab.

\section{Summary and future developments}
\label{sec:summary}

To facilitate  the future study of the role of gluons in determining the structure of the nucleon and
$J/\Psi$-nucleon ($J/\Psi$-N) interactions, we have reviewed six reaction models of 
$\gamma + p \rightarrow J/\Psi +p$ reactions.
The formulas for each model are given and used to obtain the results to show the extent to which 
the available data can be described.
The models which have been developed to describe the data from threshold to very high energy $W=300$~GeV 
are the Pomeron-exchange model of Donnachie and Landshoff ($Pom$-DL) and its extensions to include 
$J/\Psi$-N potentials extracted from LQCD ($Pom$-pot) and to also use the CQM
to account for the quark substructure of $J/\Psi$ ($Pom$-CQM).
For describing the JLab data at $ W \lesssim 7$~GeV, three models have been developed by applying  
the PQCD approach to calculate the two-gluon exchange using the generalized parton distribution of the nucleon 
($GPD$-based), two- and three-gluon exchanges using the parton distribution of the  nucleon ($2g+3g$), 
and the exchanges of scalar ($0^{++}$) and tensor ($2^{++}$) glueballs within the holographic formulation ($Holog$).
The formulas for investigating  the excitation of the  nucleon resonances $N^*(P_c)$, reported by the LHCb Collaboration,
in the  $\gamma + N \rightarrow J/\Psi +N$ reactions have also been given.
We demonstrate that the differences between these six models can be better distinguished and the $N^*$ 
can be more precisely  studied by using the forthcoming JLab data at large $-t$ and at energies very near the 
$J/\Psi$ production threshold.

We have observed that all six models are still in developing stage.
It will be useful to make the following improvements:
\begin{enumerate}
\item The $Pom$-DL and $Pom$-pot models need to replace the vector meson dominance assumption by the  
quark-loop mechanism using either the constituent quark model or the models based on Dyson-Schwinger equation of QCD. 
This is necessary to interpret the determined Pomeron-quark coupling constants.
\item $Pom$-CQM model needs to use realistic constituent quark models which fit both the charmonium spectrum and
the $J/\Psi \rightarrow e^+e^-$ decay width, and to use a parameterization of quark-nucleon potential $v_{cN}$ which 
account for the spin-dependent and momentum-dependent of the multi-gluon exchange mechanisms. 
It will be interesting to use the information from $GPD$-based model to constraint the parameters of $v_{cN}$.
It is of course essential to have the input from the most advanced LQCD calculations of $J/\Psi$-N potential, such as 
those reported in Refs.~\cite{KS13, KS15a} for the spin-dependent $J/\Psi$-N potential and Ref.~\cite{LDHI22} for the
$\phi$-N potential. The $J/\Psi$-N interaction calculated from DSE models is also highly desirable.
\item Both the $GPD$-based and $2g+3g$ models can be improved by using either CQM or DSE models to
perform quark-loop integration over the $J/\Psi$ wavefunction and to also predict $J/\Psi$-N scattering amplitudes 
which must be included to account for the final-state interaction of $\gamma + N\rightarrow J/\Psi +N$, as required  
by the unitarity condition.
\end{enumerate}

By using the reaction models with the improvements suggested above, the analysis of the forthcoming JLab data can 
help determine the $J/\Psi$-N interaction which is needed to test $J/\Psi$-N  potential extracted from lattice QCD 
and to understand the nucleon resonances $N^*(P_c)$ reported by the LHCb Collaboration. 
In addition, the determined $J/\Psi$-N potentials are also needed to investigate the production of nuclei with
hidden charms and to extract the gluon distributions in nuclei.
These goals can be accomplished by continuing the collaborations between experimental and theoretical efforts.

\begin{acknowledgements}
We are grateful to Shoichi Sasaki for providing the information on the $J/\Psi$-N potentials from LQCD
of Refs.~\cite{KS10b,sasaki-1} and to Craig Roberts for helpful discussions.
We also thank Yuxun Guo and Kiminad Mamo for their help in checking
the results from their models presented in this paper. 
The work of T.-S.H.L. was supported by the U.S. Department of Energy, Office of Science, Office of Nuclear Physics, 
under Contract No. DE-AC02-06CH11357.
The work of S.S. and Y.O. was supported by the National Research Foundation of Korea (NRF) under Grants
No. NRF-2020R1A2C1007597 and No. NRF-2018R1A6A1A06024970 (Basic Science Research Program).
\end{acknowledgements}

\begin{appendix}

\section{Regge phenomenology }
\label{sec:appendix}

There exists extensive literature~\cite{Collins,IW77,DDLN} on Regge phenomenology.
For our purposes, we will only give sufficiently self-contained explanations which are needed to present the
formulas of Pomeron-exchange  models.

\begin{figure}[t]
\centering
\includegraphics[width=0.9\columnwidth,angle=0]{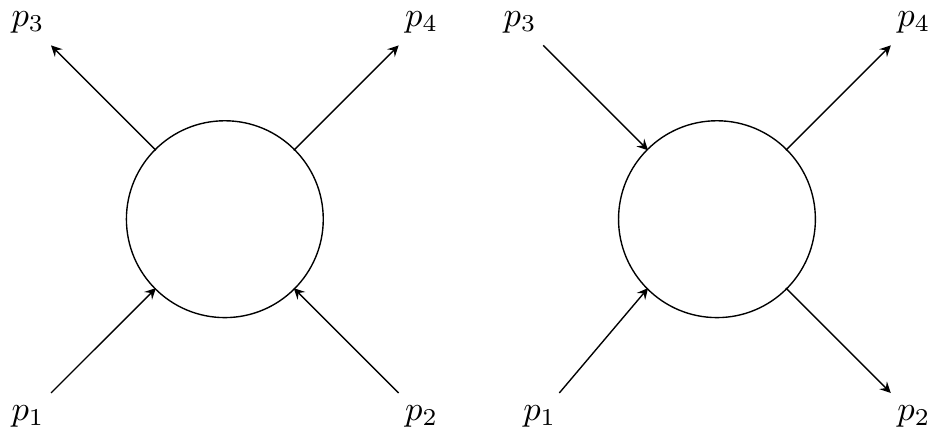}
\caption{Momentum variables of the $s$-channel (left) and $t$-channel scattering (right).}
\label{fig:st-ch}
\end{figure}

Considering the two-body scattering, the amplitudes of $s$-channel  $1(p_1)+2(p_2) \to 3(p_3)+4(p_4)$ 
and $t$-channel $1(p_1)+ 3(-p_3) \rightarrow  2(-p_2)+4(p_4)$ scattering, as shown in Fig.~\ref{fig:st-ch},  
are written in terms of the usual Mandelstam variables defined by
\begin{eqnarray}
s&=&(p_1+p_2)^2 ,\\
t&=&(p_1-p_3)^2  
\end{eqnarray}
for the $s$-channel scattering of Fig.~\ref{fig:st-ch}(left) and
\begin{eqnarray}
s_t&=&(p_1+(-p_3))^2=(p_1-p_3)^2=t , \label{eq:t-ch1} \\
t_t&=&(p_1-(-p_2))^2=(p_1+p_2)^2=s \label{eq:t-ch}
\end{eqnarray}
for the $t$-channel of Fig.~\ref{fig:st-ch}(right). 
One of the steps in developing Regge phenomenology is to assume the crossing symmetry that 
the scattering amplitudes $T(t,s)$ for the $s$-channel and $T_t(t_t,s_t)$ for the $t$-channel are related by
\begin{eqnarray}
T(t,s) &=&T_t(s,t)=T_t(t_t,s_t)\,. \label{eq:cros-1}
\end{eqnarray}
It  is important to note here that $s$ ($s_t$) is the total collision energy in the $s$-channel ($t$-channel) 
CM frame, and $t$ ($t_t$) defines the corresponding momentum-transfer of the scattering.
Thus the crossing symmetry implies  that a bound or resonance state, called  $R$, in the $t$-channel scattering
$1+3\rightarrow R \rightarrow 2+4$ can be an exchanged particle $R$ in the $s$-channel $1+2\rightarrow 3+4$ scattering.

We now describe the essential steps in getting the $s$-channel scattering amplitude $T(t,s)$ from
the $t$-channel scattering amplitude $T_t(t_t,s_t)$ by using the crossing symmetry relation of Eq.~(\ref{eq:cros-1}).
Considering $1(p_1) + 3(-p_3)  \rightarrow 2(-p_2) + 4(p_4)$ in the CM system of the $t$-channel, 
we then have the following definitions of the momentum variables:
\begin{eqnarray}
p&=&|{\bf p}_1|=|{\bf p}_3| ,\nonumber \\
q&=&|{\bf p}_2|=|{\bf p}_4| ,
\end{eqnarray}
and
\begin{eqnarray}
s_t&=&t= [E_{1}(p)+E_{3}(p)]^2=[E_2(q)+E_4(q)]^2 ,
\label{eq:s-t} \\ 
t_t&=&s=m^2_1+m^2_2+2E_1(p)E_2(q)-2\,qp \cos\theta_t , \label{eq:t-t}
\end{eqnarray}
where  $\cos\theta_t=\hat{p}_1\cdot(-\hat{p}_2)$ defines the scattering angle $\theta_t$ in $t$-channel. 
Eq.~(\ref{eq:t-t}) then leads to
\begin{eqnarray}
\cos\theta_t&=&\frac{1}{2pq} \left[ s-m^2_1-m^2_2+2E_1(p)E_2(q) \right] .
\label{eq:cos}
\end{eqnarray}
Note that $p$ and $q$ are function of $t$ as can be seen from Eq.~(\ref{eq:s-t}) and hence for a given $t$,  
$\cos\theta_t$ depends linearly on $s$.

\begin{figure}[t] 
\centering
\includegraphics[width=0.6\columnwidth,angle=0]{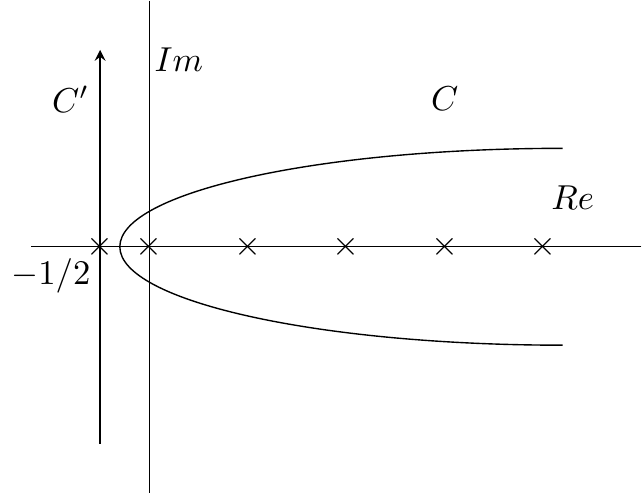}
\caption{ Contour  in l-plane.}
\label{fig:contor}
\end{figure}

The next step is to examine the partial-wave expansion of $t$-channel amplitude.
By using the relations Eqs.~(\ref{eq:s-t})-(\ref{eq:cos}), we then have
\begin{eqnarray}
T_t(t_t,s_t)
&=&T_t(t_t(\cos\,\theta_t),t)
\nonumber \\
&=&\sum_{l=0}^{\infty}(2l+1)P_l(\cos\,\theta_t) A(l,t) ,
\end{eqnarray}
where $P_l(x)$ is a Legendre polynomial in $x$. 
In the complex-$l$ plane, we apply the Watson-Sommerfeld transformation~\cite{Sommerfeld,Watson18,Watson19} 
to write the above expression as
\begin{eqnarray}
T(t,s)&=&T_t(t_t,s_t) \nonumber \\
&=&\frac{1}{2\pi i}\int_C\frac{\pi}{\sin l\pi}P_l(-\cos\theta_t)A(l,t)\, dl ,
\label{eq:con-int}
\end{eqnarray}
where $C$ is the contour indicated in Fig.~\ref{fig:contor}.
The denominator $\sin l\pi$ generates the poles (solid circles) indicated in Fig.~\ref{fig:contor}. 
Within the non-relativistic quantum mechanics, Regge~\cite{Regge59,Regge60,BLR62} showed 
that if the $s$-channel amplitude $T(t,s)$ is defined by a local potential like Yukawa potential $(\sim e^{\mu r}/r$), 
$A(l,t)$ is analytic in the complex $l$-plane, aside from poles in the $\mbox{Re}(l) \ge -1/2$. 
It can therefore be written in the following form:
\begin{eqnarray}
A(l,t)=\sum_{n}\frac{\beta_n(t)}{l-\alpha_n(t)} .
\end{eqnarray}

Closing the contour $C$ at infinity and through the $\mbox{Re}(J)=-1/2$ line, as indicated in Fig.~\ref{fig:contor}), 
and using the Cauchy's theorem, Eq.~(\ref{eq:con-int}) then becomes  
\begin{eqnarray}
T(t,s)&=&T_t(t_t,s_t) \nonumber \\
&=&\int_{-1/2-i\infty}^{-1/2+i\infty}\frac{\pi}{\sin\,l\pi}P_l(-\cos\,\theta_t)A(l,t)\, dl 
\nonumber \\ && \mbox{}
+\sum_{n}\beta_n(t)P_{\alpha_n(t)}(-\cos\theta_t)\frac{1}{\sin \pi \alpha_n(t)} .
\label{eq:Collins}
\end{eqnarray}
Here $\alpha_n(t)$ is called the Regge trajectory which leads to poles of the amplitude at
\begin{eqnarray}
\alpha_n(t= M_{L_n}^2)= L_n\,;\quad L_n = 0,1,2, \dots .
\label{eq:rpole}
\end{eqnarray}
 At these pole positions, the usual Legendre polynomial has the property $P_{\alpha_n(t)}(-\cos\theta) \rightarrow 
(-1)^{L_n}P_{L_n}(\cos\theta)$.
Thus it is suggestive that $L_n$ can be interpreted as the angular momentum of the particle formed in the 
$t$-channel process with mass $M_{L_n}$ because $t=s_t=[E_1(p)+E_2(p)]^2$.
These particles are interpreted as the exchanged particle in $s$-channel scattering. 
If this interpretation is correct, we can use the particle spectrum found in $t$-channel scattering to define 
the Regge trajectory.
Thus the main feature of the Regge phenomenology is: {\bf the particle spectrum can 
define the scattering amplitudes}.

The first term in Eq.~(\ref{eq:Collins}) is neglected in practice. 
It is also extended to define natural-parity exchange from the unnatural-parity exchange. 
The amplitude of $s$-channel scattering amplitude is then of the following form:
\begin{eqnarray}
T(s,t)&=&\sum_{n}\beta_n(t)\frac{P_{\alpha_n(t)}(-\cos\theta_t)+s_n\,P_{\alpha_n(t)}(+\cos\theta_t)}
{2\sin[\pi \alpha_n(t)]} ,
\nonumber \\
\end{eqnarray}
where the \textit{signature\/} of the trajectory, $s_n=+1$ $(-1)$ corresponds to even (odd) parity exchanges. 
In the high energy limit with very large $s$ and  $s \gg |t|$, $\cos\theta_t \sim -s/(2q(t)p(t))$, as can be seen 
from Eq.~(\ref{eq:cos}). 
It follows that
\begin{eqnarray}
P_{\alpha_n(t)}(-\cos\theta_t) \sim \left(\frac{s}{2p(t)q(t)} \right)^{\alpha_n(t)} .  
\end{eqnarray}
Here we recall Eq.~(\ref{eq:s-t}) to note that the momenta $p$ and $q$ of $t$-channel as functions of $t$ 
of the $s$-channel scattering. 
We then have
\begin{eqnarray}
T(s,t)&=&\sum_{n}F_{t}(t)
\frac{1+s_ne^{-i\pi\alpha_n(t)}}{2\sin[\pi \alpha_n(t)]} \left[ \alpha_{1,n}(t)s \right]^{\alpha_n(t)} ,
\end{eqnarray}
where
\begin{eqnarray} 
F_{t}(t)= \beta_n(t) \left(\frac{\alpha_{1,n}}{2p(t)q(t)} \right)^{-\alpha_n(t)} .
\end{eqnarray}
If we write $F_{f}(t)=\beta^{13}_n(t)\beta^{24}_n(t)$ and assume that $\beta^{13}_n(t)$ and $\beta^{24}_n(t)$ 
characterize the hadron structure, we then have the following form 
\begin{eqnarray}
T(s,t)&=&\sum_{n}\beta^{13}_n(t) \beta^{24}_n(t)
\frac{1+s_ne^{-i\pi\alpha_n(t)}}{2\sin[\pi \alpha_n(t)]} \left[ \alpha_{1,n}(t)s \right]^{\alpha_n(t)} .
\nonumber \\
\label{eq:regge-f}
\end{eqnarray}
The amplitude can then be interpreted as the exchange of particles with masses defined by $\alpha_n(t=M^2_{L_n})=  L_n$. 
This is an intuitively very attractive interpretation of the scattering. 
However, there exists no successful derivation of Eq.~(\ref{eq:regge-f}) from relativistic quantum field theory and the 
form factors $\beta^{13}_n(t)$ and $ \beta^{24}_n(t)$ are determined experimentally or calculated from a theoretical model.

\end{appendix}

\end{document}